\documentclass[preprint2,a4paper]{aastex}

\usepackage[dvipdfm]{color}

\newcommand{\footnt}[1]{\footnote{$^)$~#1}$^{)}$}

\shorttitle{Morphological classification of galaxies}
\shortauthors{Yamauchi et al.}

\begin{document}

\title{Morphological Classification of Galaxies \\
using Photometric Parameters: the Concentration index versus 
the Coarseness parameter\altaffilmark{1}
}

\author{Chisato Yamauchi\altaffilmark{2}\altaffilmark{3}, Shin-ichi Ichikawa\altaffilmark{3}, 
Mamoru Doi\altaffilmark{4}, Naoki Yasuda\altaffilmark{5},\\
Masafumi Yagi\altaffilmark{3},
Masataka Fukugita\altaffilmark{5},
Sadanori Okamura\altaffilmark{6},
Osamu Nakamura\altaffilmark{7},\\
Maki Sekiguchi\altaffilmark{5} and
Tomotsugu Goto\altaffilmark{8}\\
}

\altaffiltext{1}{
Based on the thesis by CY 
submitted to Nagoya University in 
fulfillment of MA degree requirements.
%This manuscript has been submitted to AJ, and
%will undergo \emph{major} revision after my Doctor's thesis work.
%I cannot spare enough time for this work now!.
}

\altaffiltext{2}{
Department of Physics and Astrophysics, Nagoya University, Chikusa-ku
Nagoya 464-8602, Japan: cyamauch@a.phys.nagoya-u.ac.jp}
\altaffiltext{3}{Astronomical Data Analysis Center and
Division of Optical and Infrared Astronomy,  National
Astronomical Observatory, Mitaka, Tokyo 181-8588, Japan}
\altaffiltext{4}{Institute of Astronomy, School of Science, University
of Tokyo, Mitaka, Tokyo, 181-0015, Japan}
\altaffiltext{5}{Institute for Cosmic Ray Research, University of Tokyo,
Kashiwa, 277-8582, Japan}
\altaffiltext{6}{Department of Astronomy, School of Science, University
of Tokyo, Tokyo 113-0033, Japan}
\altaffiltext{7}{School of Physics \& Astronomy University of Nottingham
 University Park Nottingham NG7 2RD, England}
\altaffiltext{8}{%Department of Physics and Astronomy, The Johns Hopkins
  %University, 3400 North Charles Street, Baltimore, MD 21218-2686, USA
  %Department of Infrared Astrophysics, 
  Institute of Space and Astronautical Science,
  Japan Aerospace Exploration Agency, 3-1-1 Yoshinodai, Sagamihara,
  Kanagawa 229-8510, Japan}

\begin{abstract}
%We improve and devise photometric parameters for the
%morphological classification using
%bright galaxies in the First Data Release
%of the Sloan Digital Sky Survey.
%In addition to using an
%elliptical-aperture concentration index for classification, we
%introduce a new texture parameter, coarseness,
%which quantifies 
%the deviation from smoothness of the isophotes.  
%Improved
%concentration index
%produces morphological classifications that are in appreciably better agreement
%with visual classifications than those based on circular aperture
%concentration indices.  
We devise improved photometric parameters for the morphological
classification of galaxies using a bright sample from the First Data
Release of the Sloan Digital Sky Survey. In addition to using an
elliptical aperture concentration index for classification, we
introduce a new texture parameter, coarseness, which quantifies
deviations from smooth galaxy isophotes. The elliptical aperture
concentration index produces morphological classifications that are in
appreciably better agreement with visual classifications than those
based on circular apertures.
With the addition of the coarseness parameter,
%the success rate of classifying galaxies into early and late types
%is increased to be 
the success rate of classifying galaxies into early and late types
increases to 
$\simeq 88$\% with respect to the reference visual classification. 
A reasonably high success rate
($\simeq 68$\%) is also attained in classifying galaxies 
into three types, early-type galaxies (E+S0), early- (Sa+Sb) and 
late- (Sc+Sdm+Im) 
type spiral galaxies.
\end{abstract}


\section{Introduction}

%Morphological classification in the Hubble sequence 
%\citep{san61}
%still serves as an important 
%quantity that represents the basic features of galaxies,
%and is probably related to their formation and evolution histories.
The morphological classification of galaxies, which assigns galaxies
%into 
discrete classes in the form of a tuning-fork diagram (the so
called Hubble sequence, Sandage 1961), allows us to quantify the basic
features of galaxies and to relate them to the galaxies' formation and
evolution histories.
While the Hubble classification is based on the 
visual inspection of images of galaxies, and 
therefore necessarily involves subjective elements,
it provides a basis for many extragalactic studies
(e.g., Dressler 1980; Binggeli, Sandage \& Tammann 1988; Dressler 1994).
%\citep{dre80,bin88,dre94}.

%With the advancement of digitized galaxy surveys,
%it is highly desirable to develop a fast, automated method of
%morphological classification
%and applicable to large quantity of data, without losing the accuracy
%of the traditional visual classification.
With the advancement of digitized galaxy surveys, it is highly
desirable to develop a fast, automated method of morphological
classification applicable to large data samples, without loosing the
accuracy of the traditional visual classification.
%Focusing on morphological classification with imaging data,
%as the title of our paper indicates,
The typical approaches employed
%by astronomers 
for morphological classification are to apply artificial neural
networks
%the two typical approaches 
%are the use of artificial neural networks 
\citep{bur92,sto92,ser93,nai95,ode02,ball04},
and characterizations with simple surface photometric 
parameters
\citep{doi93,abr94}.
In this paper, we focus on 
the photometric parameters for
the morphological
classification of galaxies using Sloan Digital Sky Survey
(SDSS, York et al. 2000; 
Early Data Release, Stoughton et al. 2002, hereafter EDR;
First Data Release, Abazajian et al. 2003, hereafter DR1;
Second Data Release, Abazajian et al. 2004)
imaging data.
%We follow the latter method, 
%since 
%%it has the advantage 
%%that the process of classification
%%is deterministic and 
%the relation between the parameters and the classification 
%can be easily understood. 
The simplest indicator often used in the
literature is the parameter that characterizes the concentration of
light towards the center of galaxies
\citep{mor58}.
%Doi et al. (1993)
%defined the concentration parameter ($C$)
%using the equivalent radii of the elliptic isophotes
%and showed
%that early- and late-type galaxies are reasonably well separated
%in isophotal photometry, with the additional use of surface brightness
%as the second parameter.
Doi et al. (1993) defined the concentration index%
%$C=r_{50}/r_{90}$%
\footnt{Strictly speaking, this is the
{\it inverse} concentration index. But 
we call it ``concentration index'' throughout this work.}
using two equivalent radii of the elliptic
isophotes. %containing 50\% and 90\% of the total galaxy light.
They show
that early- and late-type galaxies are reasonably well separated in
isophotal photometry, if surface brightness is used as a second
parameter.
\citet{abr96}
introduced a rotational asymmetry parameter
$A$ in addition to $C$ that is defined using the flux measured in 
elliptical apertures.  
%Indicator $A$ shows efficacy
%in discriminating irregular from spiral galaxies.
The rotational
asymmetry parameter allows for efficient discrimination between
irregular and spiral galaxies.
The $\log C$-$\log A$ diagram has been used as a 
tool to classify morphology of distant galaxies 
observed with the HST
%taken with HST
(e.g., Abraham et al. 1996; Brinchmann et al. 1998).
\citet{tak99} discusses the evolution of the structure of galaxies
using $\chi$ parameter calculated from the residual image,
which indicates the power at high spatial frequencies in the disk of 
the galaxy. 
\citet{shi01} and \citet{str01} showed that 
%the (inverse) concentration index $C$(=$r_{50}/r_{90}$) defined using
%Petrosian circular apertures, 
%computed and cataloged in the SDSS
%%Sloan Digital Sky Survey 
%%(SDSS, York et al. 2000; 
%%Early Data Release, Stoughton et al. 2002, hereafter EDR;
%%First Data Release, Abazajian et al. 2003, hereafter DR1;
%%Second Data Release, Abazajian et al. 2004),
%correlates fairly well with the morphological type and can be used for
%a classification into early- and late-type galaxies.
the concentration index, calculated using circular
isophotes as part of the standard SDSS pipeline reductions (Stoughton
et al. 2002), correlates fairly well with morphological type and can
be used for classification into early and late galaxy types.
%better than with the use of other simple photometric parameters.
The success rate of these approaches, with reference to visual inspection,
is approximately 
%$\approx 80$\%.
$80$\%.

In this paper, we use the SDSS Data Release 1 (DR1, Abazajian et
al. 2004) imaging data of 1421 galaxies selected from the SDSS Early Data
Release (EDR, Sthoughton et al. 2002) 
and early commissioning data
to improve the success rate of
morphological classifiers by introducing additional photometric
parameters. First, we consider the performance of the standard
circular-aperture concentration index, with particular emphasis on
the cases where the classification fails. We proceed to employ the
elliptical definition of Doi et al. (1993) and Abraham et al. (1996)
for the measurement of isophotal and aperture fluxes and compare the
success rate to that obtained using the circular definition. We also
introduce a new texture parameter to help our classification. The
motivation for this approach is that,
in the visual classification, we mainly resort to surface brightness texture,
such as properties of spiral arms, clumpiness, 
and  HII regions, in addition to
the concentration of light towards the galaxy center. 
The only work 
that we are aware of that discusses texture parameters as a classifier is that
by \citet{nai97}.
They proposed
``blobbiness'', ``isophotal center displacement'' and ``skeleton
ratio'' parameters, which correspond to roughness, global asymmetries and 
more localized structure, respectively. 
These parameters %certainly 
correlate with galaxy morphology and may be useful for classification, but our
own tests show that they do not enhance the success rate of the
classifier. Our texture parameter -- coarseness -- describes the
structure of the outer galaxy isophotes, emulating the method employed
in visual classification. 
%The flux fluctuations are computed along the elliptic
%circumference; this characterizes the departure from a smooth 
%surface profile of the galaxy.
The flux fluctuations are computed along elliptical circumferences in
order to characterize the galaxy profile's departure from smoothness.
The coarseness parameter is defined as the
ratio of the range of
fluctuations in surface brightness 
along the elliptic circumference to the full dynamic range of surface
brightness
of the galaxy. In this study we evaluate the
performance of the elliptical concentration index and the coarseness
parameter as classification parameters (used separately or together)
in comparison to the visually classified SDSS sample of Fukugita et
al. (in preparation).

This paper is organized as follows. The
elliptical-aperture concentration index is defined 
%in Section 2.
%The concentration index using elliptical apertures is defined
in section \ref{Ce}.
The coarseness parameter is introduced  
in section \ref{Y}.
After briefly describing our sample in section \ref{sample}, the
correlation of the two parameters with the visually obtained morphology
is investigated in section \ref{behavior_of_Ce}. 
We study the performance of the 
morphological classification using these 
parameters in section \ref{classification},
and conclusions are presented in section \ref{conclusions}.

\section{CONCENTRATION INDEX}
\label{Ce}

%The SDSS uses the Petrosian flux to define photometric quantities.
We consider a concentration index for elliptical apertures using
Petrosian quantities. 
%We define the intensity weighted second order moments by
%%We adopt the standard definitions implemented in the
%%SDSS photometric pipeline \emph{Photo} ...
The intensity-weighted second-order moments
are defined 
in the SDSS photometric pipeline 
(hereafter PHOTO; Lupton 1996; Lupton et al. 2001) 
as
\begin{equation}
\label{eq_moments}
M_{xx} \equiv \langle x^2 / r^2 \rangle,~
M_{xy} \equiv \langle xy / r^2 \rangle,~
M_{yy} \equiv \langle y^2 / r^2 \rangle .
\end{equation} 
If the major and minor axes of the ellipse lie along the
$x$ and $y$ axes, the axis ratio $\alpha = b/a$ is calculated as
\begin{equation}
\alpha = \frac{M_{yy}}{M_{xx}}.
\end{equation}
because
\begin{equation}
\label{eq_mxxmxymyy}
M_{xx} = \frac{1}{1+\alpha},~~~~
M_{xy} = 0,~~~~
M_{yy} = \frac{\alpha}{1+\alpha}.
\end{equation}

In general, we rotate the image to account for the position angle $\phi$
from $(x',y')$ to $(x,y)$ to align the galaxy image 
along the axes:
\begin{equation}
 { x' = x \cos \phi - y \sin \phi \atop 
 y' = x \sin \phi + y \cos \phi .}
\label{eq:rot}
\end{equation}
The Stokes parameters $U$ and $Q$ are calculated as 
\begin{eqnarray}
%\begin{equation}
\label{mxy_p}
 U/2\equiv M_{x'y'} &=& (M_{xx}-M_{yy}) \sin \phi \cos \phi \nonumber\\
  &=&    \frac{1-\alpha}{1+\alpha} \cdot \frac{\sin 2 \phi}{2} 
%\end{equation}
\end{eqnarray}
\begin{equation}
\label{mxx_p_xyy_p}
 Q\equiv M_{x'x'} - M_{y'y'} = \frac{1-\alpha}{1+\alpha}\cos 2 \phi,
\end{equation}   
where $\phi=1/2~{\rm tan}^{-1}(U/Q)$. The axis ratio is $\alpha=(1+P)/(1-P)$
with $P=\sqrt{(U^2+Q^2)}$. $U$ and $Q$ are evaluated in 
%Photometric Pipeline 
PHOTO
and cataloged in the SDSS data releases.
%(Lupton 1996; Lupton et al. 2001).
%in the SDSS photometric pipeline 
%(hereafter PHOTO; Lupton et al. 1996; 2001) 

To calculate the concentration index $C_e$ for the ellipse,
we consider the area $A_e(a)$ of the ellipse of the semi-major axis
$a$ and axis ratio $\alpha$, and the integrated flux $F_e(a)$ within $A_e(a)$.
We define the Petrosian semi-major axis $a_{\rm P}$ for a given $\eta$ as 
\begin{equation}
\scriptsize
\eta =
 \frac{ \{ F_e(1.25a_{\rm P})-F_e(0.8a_{\rm P}) \} / \{ A_e(1.25a_{\rm P})-A_e(0.8a_{\rm P}) \} }
      { F_e(a_{\rm P}) / A_e(a_{\rm P}) }, 
\end{equation}
where we take $\eta$=$0.2$, and
the elliptical Petrosian flux $F_{\rm P}$ as
\begin{equation}
F_{\rm P} = F_e(ka_{\rm P})
\end{equation}
with $k$ set equal to 2, following the SDSS definition
\citep{str02}.

The Petrosian half-light and 90\%-light
semi-major axes $a_{\rm 50}$, $a_{\rm 90}$ are defined 
in such a way that the flux in the elliptical apertures of these
semi-major axes is
50\% and 90\% of the elliptical Petrosian flux:
\begin{equation}
 F_e(a_{50}) = 0.5F_{\rm P},~~F_e(a_{90})=0.9F_{\rm P}~.
\label{eq:F_p}
\end{equation}
We define our concentration index $C_{e}$ 
by 
\begin{equation}
 C_{e} = a_{\rm 50} / a_{\rm 90}~.
\end{equation}

\section{THE COARSENESS PARAMETER}
\label{Y}

\begin{figure}[!b]
 \begin{center}
 \includegraphics[scale=0.56]{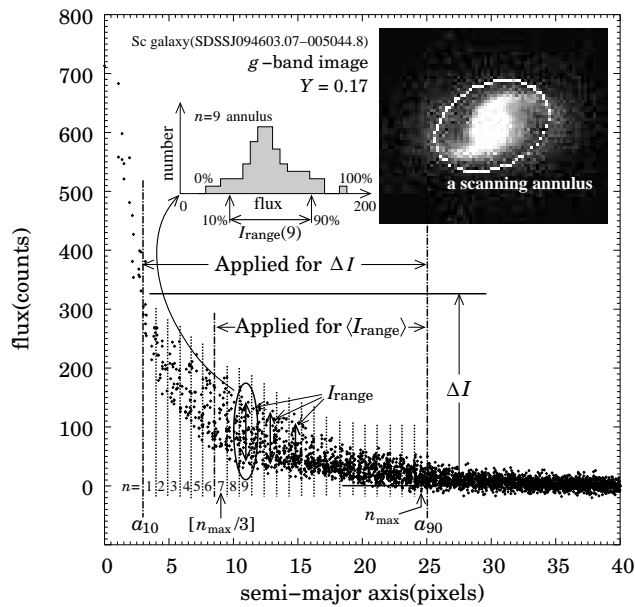}
 \vspace*{-7mm}
 \end{center}
 \caption{An illustration
 explaining the method of computing $Y$ for the galaxy image inlaid in
the top right of the figure. The figure shows 
%the way dividing 
division of
the area sandwiched by the two ellipses
specified by $a_{90}$ and $a_{10}$ into $n_{\rm max}$
annuli. 
The dotted lines show division into elliptical annuli.
The range of flux $I_{\rm range}(n)$ is defined by 
the 90\%- and 10\%-tiles of the flux distribution of $n$-th annulus
as an inlaid histogram in the top left of the figure.
%as $I_{90}(n)$ and $I_{10}(n)$, respectively().
The $I_{\rm range}$ computed in each annulus
is averaged over $N$ annuli to give $\langle I_{\rm range} \rangle$, 
and the coarseness parameter
$Y$ is given as $\langle I_{\rm range} 
 \rangle$ divided by the full dynamic range of the flux observed
for the galaxy, $\Delta I$.
}
 \label{example_profile}
\end{figure}
% da, dec = 146.512825, -0.845782 

\subsection{Definition of a Texture Parameter}

In this section we provide a detailed ``step-by-step'' method for the
calculation of the coarseness parameter.
The position angle $\phi$ and the axis ratio
$\alpha$ are calculated in the same way as in the 
previous section.
Each position $(x,y)$ of pixel on the image
is calculated by equation (\ref{eq:rot}) beforehand, so that
the semimajor and semiminor axes are aligned with the $x$ and $y$ axis.

We consider successive elliptical annuli from $a = a_{\rm 10}$ outwards to 
$a_{\rm 90}$, where $a_f$ is defined by $F_e(a_{f})=\frac{f}{100}F_{\rm P}$
as in equation (\ref{eq:F_p}),
assuming that 
each elliptical annulus has the same position angle and
is congruent.
We define the `equivalent distance' $d$ from the center of the galaxy
to a pixel at $(x,y)$
%, after rotation of the original position
%by (\ref{eq:rot}), 
by
\begin{equation}
 d = \sqrt{ x^2 + (y/\alpha)^2 }.
\end{equation}
We then divide the area sandwiched by the two ellipses specified 
by $a_{\rm 10}$ and $a_{\rm 90}$
into $n_{\rm max}$ annuli, as specified in what follows:
We calculate $d$ of each pixel contained in the annulus between 
$a_{\rm 10}$ and $a_{\rm 90}$, and
%range the pixels in the ascending order with respect to $d$ as
%% thank you, Dave!!
place the pixels in ascending order with respect to $d$ as
\begin{equation}
\scriptsize
 \underbrace{  d_{1_1},d_{1_2}, \cdot\cdot\cdot,d_{1_{{\cal N}_{1}}}  },~
 \underbrace{  d_{2_1},d_{2_2}, \cdot\cdot\cdot,d_{2_{{\cal N}_{2}}}  },~
 \underbrace{  d_{3_1},d_{3_2}, \cdot\cdot\cdot,d_{3_{{\cal N}_{3}}}  },~
 \cdot\cdot\cdot~, ~ \atop
 {\rm 1st~annulus~~~~~~~~~~ 2nd~annulus~~~~~~~~~~ 3rd~annulus}~~
 \cdot\cdot\cdot ~~~~
\label{eq:d1d2d3}
\end{equation}
where the number of pixels contained in the $n$-th annulus ${\cal N}_{n}$ 
is
%fixed
calculated
by the equivalent distance to the innermost point in
the $n$-th annulus, $d_{n_1}$:
%, in the way that
%That is, the ${\cal N}_{n}$ is defined by

\begin{equation}
 {\cal N}_{n} = [ 2 \pi  \alpha d_{n_1} ]~,
\label{eq:N-n}
\end{equation}
%which specifies the division into annuli,
%where $\alpha$ is the axis ratio defined in the previous section.
where $[x]$ is the integer part of $x$.
The last annulus, which terminates
at $a_{90}$, does not generally satisfy the condition (\ref{eq:N-n}). 
We show the actual algorithm which satisfies equation 
(\ref{eq:d1d2d3}) and (\ref{eq:N-n}) as follows:

  1) Focus on 1st $d$, $d_{1_1}$.

  2) Calculate ${\cal N}_1$ by equation (\ref{eq:N-n}), 
     ${\cal N}_1 = [ 2 \pi  \alpha d_{1_1} ]$.

  3) When ${\cal N}_1$=6 for an example%
\footnt{%
%If $d_{1_6}$ is equal to $d_{1_{6+x}}$, ${\cal N}_1$ is set to $6+x$.
If equidistant pixels are present, e.g., $d_{1_6}$is equal to $d_{1_{6+x}}$ for
some integer $x$, ${\cal N}_1$ is set to $6+x$.
}%
, the member of 1st annulus is
\[
     d_{1_1}, d_{1_2}, d_{1_3}, d_{1_4}, d_{1_5} ~  {\rm and} ~ d_{1_6}.
\]

  4) $d_{2_1}$ is automatically determined by $d_{1_6}$
     ($d_{2_1}$ is next to $d_{1_6}$. That is, 7th $d$).

  5) Focus on 7th $d$, $d_{2_1}$.

  6) Calculate ${\cal N}_2$ by equation (\ref{eq:N-n}), 
     ${\cal N}_2 = [ 2 \pi  \alpha d_{2_1} ]$ $\cdot\cdot\cdot$

Let us take the $n$-th annulus, and
consider the flux distribution. 
We denote the 90\%- and 10\%-tiles of the flux distribution
in the $n$-th annulus as $I_{90}(n)$ and $I_{10}(n)$, respectively
(see the inlaid histogram in the top left of Figure
\ref{example_profile}),
and the range of the two values as 
\begin{equation}
 I_{\rm range}(n) = I_{90}(n) - I_{10}(n) ~.
\end{equation}
We define the mean of $I_{\rm range }(n)$ over the annuli between 
$n_{\rm begin}$ and $n_{\rm max}$, 
as
\begin{equation}
 \langle I_{\rm range } \rangle = 
  \frac{1}{n_{\rm max}-(n_{\rm begin}-1)}
  \sum_{n=n_{\rm begin}}^{n_{\rm max}} I_{\rm range}(n),
\end{equation}
where the mean is taken 
 from the $n_{\rm begin}$-th to the outermost annulus,
some inner annuli being excluded. 
%This exclusion of the inner
%annuli is made to 
The exclusion of the inner radius is necessary in order to
enhance the visibility of texture, which
is usually more pronounced in outer regions for late-type galaxies.
From trial and error, we adopted $n_{\rm begin}=[n_{\rm max}/3]$ with
the floor function
%greatest integer function
as the best value for the performance of classification.

To further enhance the signal, we subtract the contribution of the sky noise from
$\langle I_{\rm range } \rangle$,
\begin{equation}
I_{\rm signal}= \sqrt{
  \langle I_{\rm range } \rangle^2 - (2.56 \cdot \varsigma_{\rm sky})^2}\ ,
\label{eq:i-signal}
\end{equation}
where $\varsigma_{\rm sky}$ denotes the rms of the sky noise.
We multiply the sky noise by a factor of 2.56 
so that its strength corresponds to the range between
90 and 10\%-tiles in the Gaussian distribution to match our
definition of $I_{\rm range }$.
With this choice, $I_{\rm signal}$ vanishes
when the frame does not contain objects, i.e., contains 
sky noise only.
%This subtraction enhances the performance of the classifier.
%We find that this subtraction enhances
%the performance. 

We then divide $I_{\rm signal}$ by the dynamic range of the image 
$\Delta I$, i.e.,
\begin{equation}
 Y = \frac{ I_{\rm signal}}{\Delta I}~.
\label{eq_of_Y}
\end{equation}
The $\Delta I$ is defined by
\begin{equation}
 \Delta I = {\rm max} \{I_{90}(n)\} - {\rm min} \{I_{10}(n)\}~,
\end{equation}
where ${\rm max}\{\}$ and ${\rm min}\{\}$
are taken from $ 1 \le n \le n_{\rm max}$.
%Note that we use all annuli for $\Delta I$, not from 
%$n_{\rm begin} \le n \le n_{\rm max}$. This is the result from trial and error.
We use all annuli in the computation of $\Delta I$, not just
$n_{\rm begin} \le n \le n_{\rm max}$,
which allows us to increase the signal contrast.

This procedure is sketched in Figure \ref{example_profile},
%The data show the flux of a test image (inlaid in the figure) 
%detected in each pixel, which is plotted as a 
%function of semi-major axis of the ellipse.
which shows the flux of each
pixel of a test image plotted as a function of its semi-major axis.
%
%Note that the $Y$ parameter vanishes if the profile is %an ideal,
%smooth function such as a model de Vaucouleurs or a model exponential 
%profile (in the absence of noise), since in such case $I_{\rm range}=0$.
%When the profile exhibits structures such as
%arms, both $I_{\rm range}$ and $Y$ become non-zero.
%
Note that, in the absence of noise, the $Y$ parameter vanishes if the
profile is a smooth function (such as a model de Vaucouleurs or a
model exponential profile), since this corresponds to $I_{\rm range}=0$. On
the other hand, if structures such as spiral arms are present, both
$I_{\rm range}$ and $Y$ become non-zero.

\subsection{Image Rescaling}

\begin{figure}[!b]
 \begin{center}
  \includegraphics[scale=0.35]{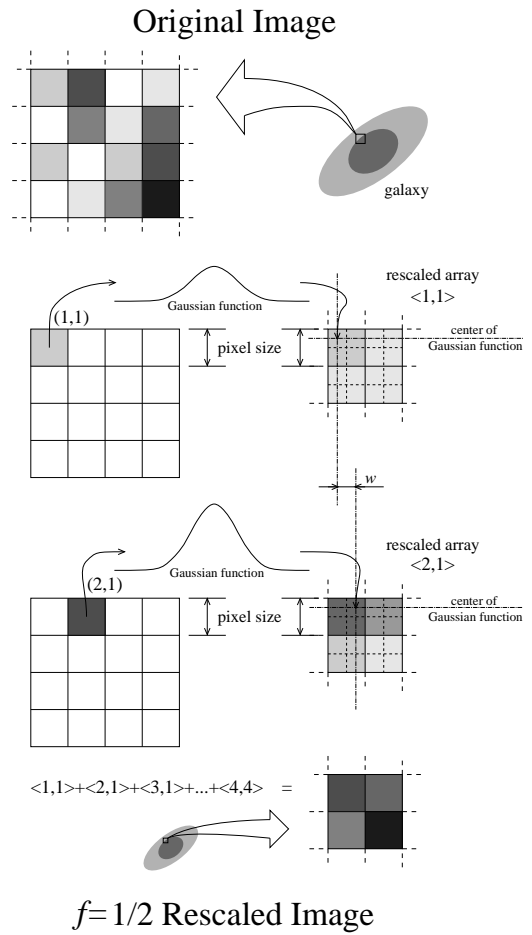}
 \vspace*{-3mm}
 \end{center}
 \caption{The method of creating rescaled images ($f=1/2$).
 Each Gaussian function whose intensity is $I(x,y)\cdot f^2$
 and width is adjusted to the effective seeing 
 will be mapped to the rescaled array. 
 The mapping $<$1,1$>$+$<$2,1$>$+...+$<$4,4$>$ which creates
 the rescaled image conserves the surface brightness.
 }
 \label{gaussian_filter}
\end{figure}

The coarseness parameter thus defined may  
depend on the apparent size of galaxies, because larger images 
are resolved in finer details. To avoid this dependence,
we set the reference size of the galaxy image, and reduce the size 
of larger galaxies to the reference size. 
The coarseness parameter is measured for the rescaled image.
The number of pixels in the rescaled image is taken to be that 
in the reference image.
%
%The reduction of image size takes 
%into account the fact that the rescaling factor is generally 
%non-integer and the seeing depends on the individual images, as we
%explain in what follows (see also Figure \ref{gaussian_filter}
%for illustration). 
The reduction of image size takes into account both the fact that the
rescaling factor is generally non-integer and the image variations in
seeing, as explained below and illustrated in Figure \ref{gaussian_filter}.
We exclude from further consideration galaxies with
apparent sizes smaller than our chosen reference size.

We rescale the image with the semimajor axis $a_{\rm 90}$ to
the reference size image with $a_{\rm 90}^{\rm ref}$ using 
Gaussian functions. We assign a Gaussian function to each pixel 
and map it to the rescaled array. The Gaussian function for the
pixel at $(x,y)$ 
%is constructed so that its integration is  $I(x,y)f^2$ where
is constructed so that its integral equals $I(x,y)f^2$,
with the rescaling factor, $f$, given by:
\begin{equation}
 f =\frac{a_{90}^{\rm ref}}{a_{90}}~.
\end{equation}
%is the rescaling factor and the Gaussian width $\sigma_{xy}$,
The Gaussian width $\sigma_{xy}$ is
\begin{equation}
 \sigma_{xy} = \sqrt{ \sigma_{\rm ref}^2 - (\sigma_{\rm
 seeing} f)^2 } ~.
\label{eq_of_root}
\end{equation} 
where $\sigma_{\rm ref}$ is the seeing taken as our reference and
$\sigma_{\rm seeing}$ is the actual seeing of each image.
With this procedure the PSF in the rescaled image is standardized
by varying $\sigma_{xy}$ from image to image.
The Gaussian functions with $\sigma_{xy}$ 
are 
%mapped
projected onto a new array with the 
interval of pixels
\begin{equation}
 w = f \times {\rm pixel~size}~.
\end{equation}
This projection conserves surface brightness.

\begin{figure}[!t]
 \begin{center}
  \includegraphics[scale=0.55]{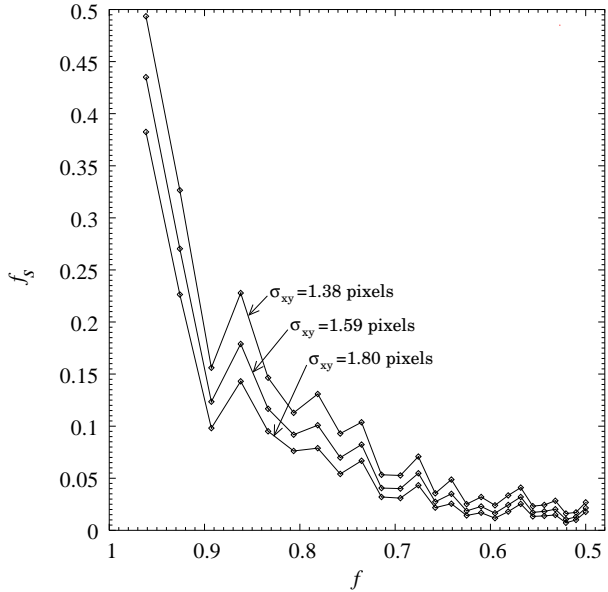}
 \vspace*{-5mm}
 \end{center}
 \caption{Effective ratio of sky noise $f_s$ for the rescaled image
as a function of the
% original sky noise $f$ 
scaling factor $f$ for a given seeing $\sigma_{xy}$. 
 }
 \label{f-fs_relation}
\end{figure}

%The sky noise that appears in (\ref{eq:i-signal}) after rescaling the image
%differs from the observed value.
%We must rescale the sky noise as 
The use of rescaled images requires similar adjustment of the observed sky
noise $\varsigma$:
\begin{equation}
 \varsigma_{\rm sky} = f_s \cdot \varsigma,
\end{equation}
%where $f_s$ is calculated by creating an artificial sky 
%of Gaussian noise,
%applying the same Gaussian smearing,
%and measuring the standard deviation ($f_s$).
%The quantity $f_s$ is dependent on $\sigma_{xy}$ and the pixel size.
%The $f_s$ thus calculated is
%shown in Figure \ref{f-fs_relation}. 
where $f_s$ is the effective ratio 
%standard deviation 
of standard deviation obtained from an artificial sky-noise image
to which equivalent Gaussian smearing has been applied. The quantity
$f_s$, shown in Figure \ref{f-fs_relation},
depends on both $\sigma_{xy}$ and the pixel size.

\begin{table}[!b]
\begin{center}
 \scriptsize
 \begin{tabular}{ccrrrr}
\hline\hline
             &~~ & \multicolumn{2}{c}{Total Sample} & \multicolumn{2}{c}{Our Sample} \\
 Hubble Type & $T$  & Number & Percent & Number & Ratio\\
 \hline 
 unclassified & $-$1 & 23  & 1.3\%   & ---  & --- \\
 E  & 0            &  242  & 13.3\%  & 194  & 13.7\%\\
 S0 & 0.5, 1       &  494  & 27.2\%  & 343  & 24.1\% \\
 Sa & 1.5, 2       &  309  & 17.0\%  & 211  & 14.8\%\\
 Sb & 2.5, 3       &  301  & 16.6\%  & 250  & 17.6\%\\
 Sc & 3.5, 4       &  337  & 18.5\%  & 314  & 22.1\%\\
 Sdm & 4.5, 5      &  80   & 4.4\%   & 80   & 5.6\%\\
 Im  & 5.5, 6      &  31   & 1.7\%   & 29   & 2.0\%\\
 \hline 
 Total &           & 1817  & 100\%   & 1421 & 100\%\\
\hline
 \end{tabular}
\end{center}
 \caption{Morphological compositions of the total sample 
  and the sample after applying the size cutoff and
  dropping $T$=$-1$ class.}
 \label{histo_t}
\end{table}

\section{THE SAMPLE}
\label{sample}

The galaxies used in our test are 
taken from the northern equatorial stripe given in SDSS DR1
\citep{aba03}.
The photometric system, imaging hardware and 
astrometric calibration of SDSS
are described in detail elsewhere
\citep{fuk96,gun98,hog01,smi02,pie03}.
We use the catalog of visual morphological classification
%the visually classified SDSS sample of 
%carried out by 
%due to
provided by
Fukugita et al. (in preparation; 
see also Nakamura et al. 2003) based on the $g$-band image. 
%% insert %%
This catalog is based on the SDSS-EDR and early commissioning data,
and contains 1875 galaxies brighter
than $r^{*}$=$15.9$, where $r^{*}$ is the
extinction corrected Petrosian magnitude
\footnt{%
The notation $r^{*}$ denotes the preliminary nature of the early
photometric calibration of SDSS commissioning data used for the
original sample selection (see Shimasaku et al. 2001).
}.
%We performed the positional matching to identify galaxies in the
%Fukugita et al. catalog
%with those in DR1, since the deblending of EDR and 
%early commissioning data presents a problem for large galaxies.
%
%The positional matching prevents the appearance of non-real
%galaxies due to the deblend of larger galaxies,
%and then the total sample
%was reduced to 1817 galaxies with dropping non-real 58 galaxies.
%
%Fukugita et al. gave $T$=$-1$ (unclassified) for
%morphologically disturbed 23 galaxies in 1817;
%we do not use these galaxies for our analysis.
%
Using positional matching we identify galaxies in the Fukugita et
al. catalog with DR1 objects. This allows us to weed out
``fake'' galaxies present in the catalog as a result of erroneous galaxy
deblends in the EDR and early commissioning data. This positional
matching reduces the number of galaxies in the Fukugita et al. catalog
by 58, and then 
the number of the total sample is reduced to 1817 galaxies.
The galaxies are classified into $T$=$-1$ as unclassified and
13 morphological types from 
$T$=$0$ (corresponding to E in the Hubble type) to $6$ (Im)
allowing for half integer classes. In this paper, we mostly refer
to $T$=0 as E, 0.5 and 1 as S0, 1.5 and 2 as Sa, ..., 5.5
and 6 as Im,
as designated in Table \ref{histo_t}, where the number of
galaxies in each class is shown. 

\begin{figure}[!b]
 \begin{center}
 \includegraphics[scale=0.25]{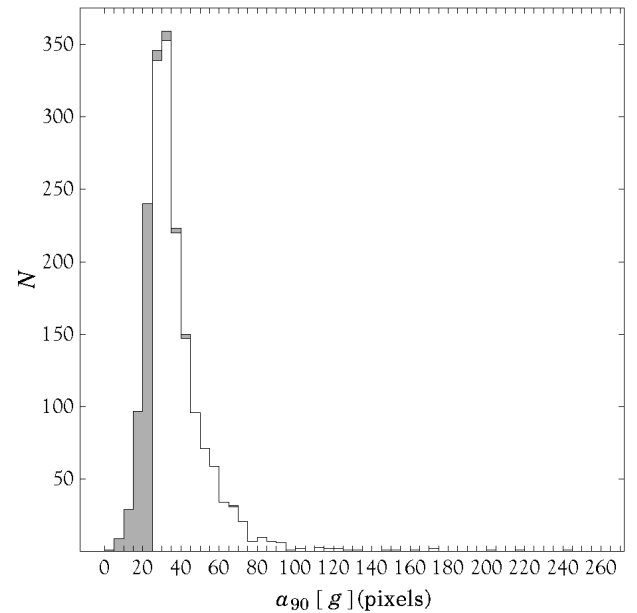}
 \vspace*{-7mm}
 \end{center}
 \caption{%
 Distribution of the semi-major axes
 containing 90\% of the Petrosian flux of the 1817 galaxies in the SDSS
 $g$-band (1 pixel = 0.4 arcsec).
 %Distribution of the Petrosian 90\% semi-major axis of the
 %$g$-band images (1 pixel = 0.40 arcsec) for 1817 galaxies.
 %The region shown by shading shows galaxies removed from our
 %consideration, leaving 1421 galaxies for our sample.
 The shaded region indicates galaxies with
 small angular sizes ($a_{90}<25 $pixels)
 and unclassified galaxies which were removed from
 consideration (396 in total).
 }
 \label{histo_re90g}
\end{figure}

\begin{figure}[!b]
 \begin{center}
 \includegraphics[scale=0.25]{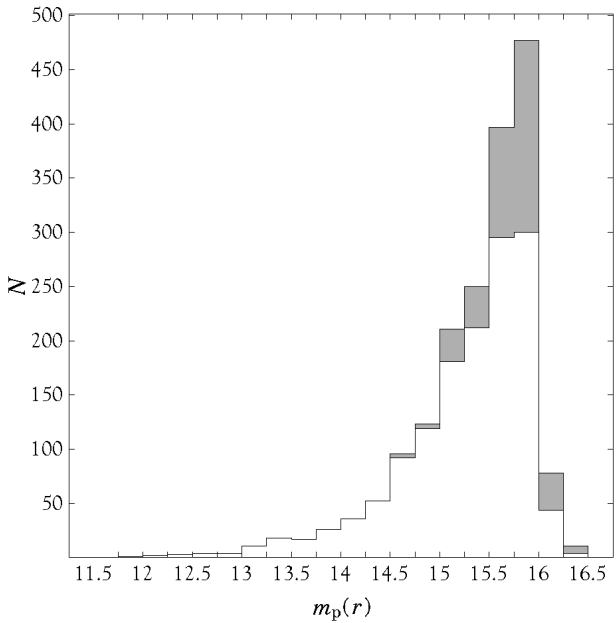}
 \vspace*{-7mm}
 \end{center}
 \caption{Distribution of the $r$-band Petrosian magnitude of the
  1817 galaxies. 
 The shaded region indicates galaxies with
 small angular sizes ($a_{90}<25 $pixels)
 and unclassified galaxies which were removed from
 consideration (396 in total).
 %The region shown by shading shows galaxies removed from our
 %consideration.
 }
 \label{histo_mpg}
\end{figure}

The $r$-band image is used to compute the concentration index.
We use the original image without rescaling, since the seeing effects
on the concentration index are small for our bright galaxy sample.
The coarseness parameter is computed 
using the $g$-band image because 
this filter is generally thought to be more sensitive to
texture than $r$-band data (Our experiments have shown, however, that
the use of the $r$-band image leads to
very little difference in classification).
%The reference size $a_{\rm 90}^{\rm ref}$ was set
%%to be
%at
%$a_{\rm 90}^{\rm ref}$ to be 25 pixels %(=$10''$)
%\citep[=$10''$;][]{gun98}
%as a test;
%we discarded galaxies with $a_{\rm 90}$
%smaller than this size. 
We set the reference size of the
semimajor axis to $a_{\rm 90}^{\rm ref}$ = 25 pixels 
\citep[=$10''$;][]{gun98}
and
exclude all galaxies with smaller semimajor axes from consideration.
This selection for the reference size of the semimajor axis was found
to be appropriate for the computation of the coarseness parameter, as
galaxies of larger sizes rescaled to 25 pixels retain enough detail
for effective classification.
In addition, we exclude 23 morphologically disturbed galaxies
of class $T$=$-1$ (unclassified by Fukugita et al.) from further
analysis.
Figure \ref{histo_re90g} shows the galaxy size distribution
measured in the $g$-band,
and the shaded areas represent the objects eliminated by the size cutoff
and $T$=$-1$ class.
This selection leaves the 1421 galaxies
in our sample, primarily removing 
galaxies of earlier types. 
Galaxies of later types (including
irregular galaxies) generally have larger size in a magnitude-limited
sample, and are little affected by the size cutoff (see Table  \ref{histo_t}).
%How this selection affects the flux (in the $r$-band)
%limited sample is displayed in Figure \ref{histo_mpg}.
The effect of this size cutoff on the $r$-band limited sample is
displayed in Figure \ref{histo_mpg}.
This selection certainly causes a bias in the ratio
among the number of morphological types, but we are not concerned in this paper
with issues of completeness and statistics of
the number distribution of morphology.

We fix FWHM of the seeing at 3.53 pixels
($1.40''$; the median of our sample) and 
the observed sky noise
$\varsigma$ at 4.0
counts for rescaling
to simplify our analysis.
Since our sample contains bright galaxies whose
$a_{\rm 90}$ is near $a_{\rm 90}^{\rm ref}$ and variations of the seeing and
$\varsigma$ are sufficiently small, 
%the simplicity
this simplification
does not seriously
affect the results.
%(Meanwhile, too large $a_{\rm 90}$ of some of our
%galaxies affects the results of the coarseness parameter according to our
%investigation, so we rescale the images).  
%For a more elaborate analysis or for galaxies of small sizes or fainter
%galaxies, one should take account of variations of these quantities from
%image to image.
In order to assure the independence of the coarseness parameter on
galaxy size, we rescale all images to $a_{\rm 90}^{\rm ref}$=25. Extending the
analysis to smaller and fainter galaxies will require proper account
of the image-to-image variations in seeing and sky noise.

\section{PHOTOMETRIC PARAMETERS AND THE MORPHOLOGICAL TYPE}
\label{behavior_of_Ce}

%In this section, we present how 
%two photometric parameters behave when
%they are plotted against the visual morphological type,
Here we investigate the behavior of the three photometric parameters
(circular and elliptical concentration index and coarseness) as a
function of the visual morphological type,
as quantified by the morphological index $T$.
%using the SDSS sample in section \ref{sample}.
We also show the effect of the axis ratio on these parameters.

\subsection{Concentration parameters}

\begin{figure}[!b]
 \begin{center}
 \includegraphics[scale=0.255]{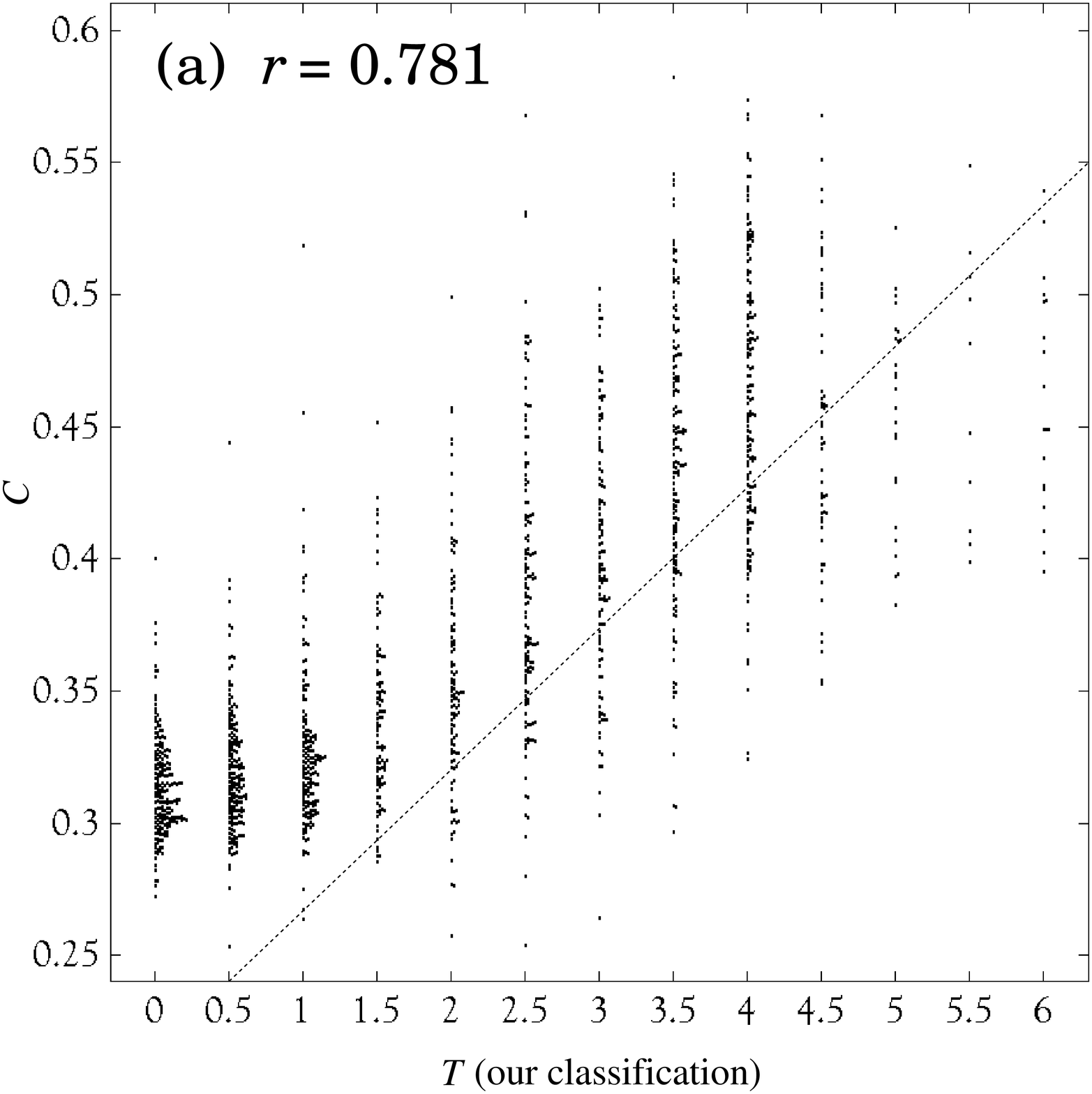} \\
% \vspace*{-1mm}
 \includegraphics[scale=0.255]{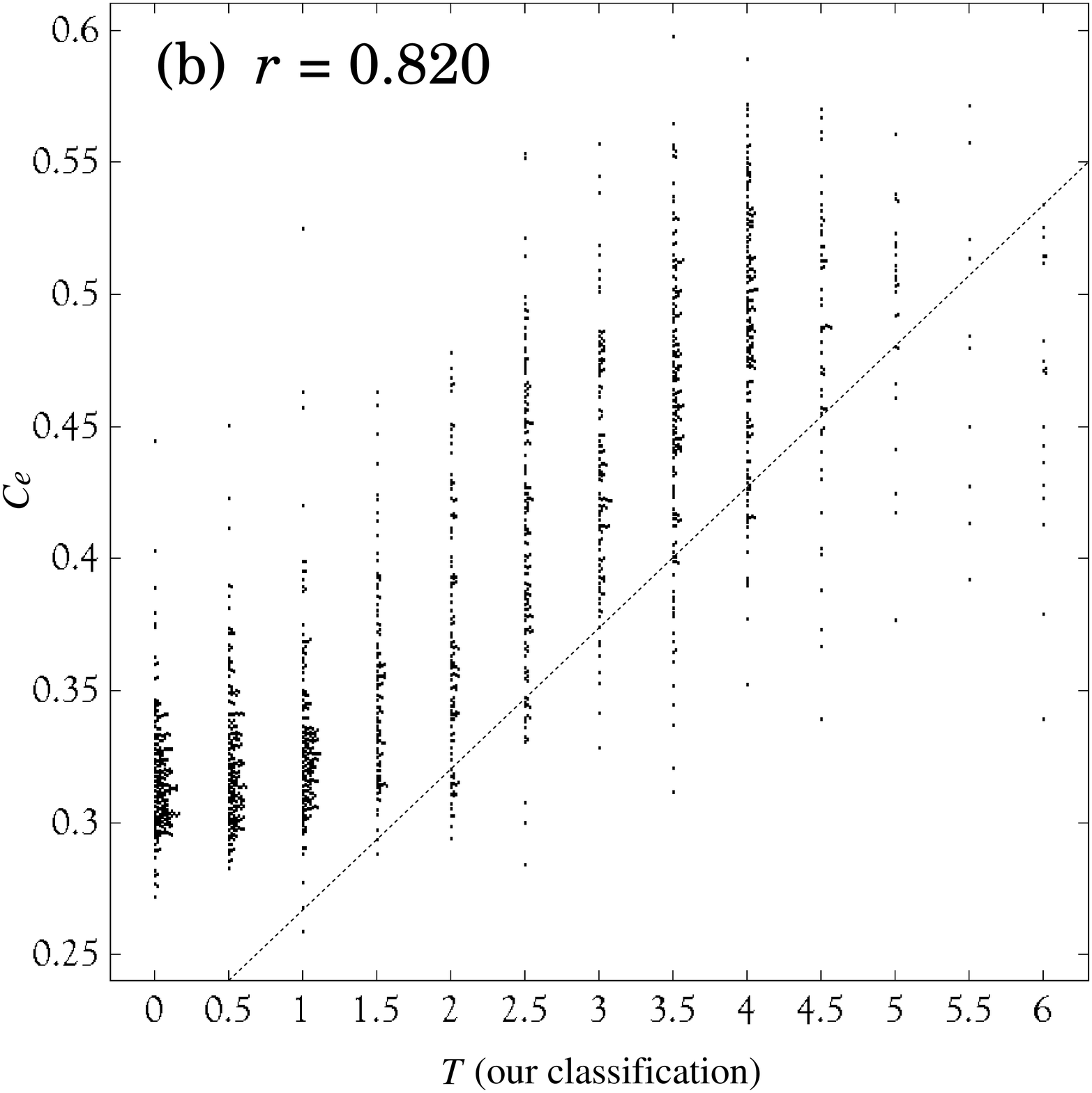} \\
% \vspace*{-5mm}
 \end{center}
 \caption{%
%Inverse concentration indices of 1421 galaxies 
% plotted against morphology $T$: (a) with the standard 
% circular apertures used in the SDSS DR1 catalog; (b) 
% with elliptical apertures defined in the text.
% To show the distribution of the points more clearly, 
% some of the dots are shifted to the right, if needed,
% to avoid overlaps.
% The number on the top of each panel is a linear correlation coefficient.
% The dotted lines are arbitrarily placed in the same position 
% to show
% differences under the lines between (a) and (b).
The concentration
indices of the 1421 galaxies in our sample versus the visual morphology
index $T$. In panel (a) we use the standard circular aperture
definitions from the SDSS DR1 catalog for computing the concentration
index; in panel (b) we use elliptical apertures. Some points were
shifted to larger $T$ values to avoid overlaps. The linear correlation
coefficient is given in the top left corner of each panel. The dotted
line is an arbitrary line placed in the same position in both panels to
facilitate comparison.
}
 \label{t_vs_cin}
\end{figure}

\begin{figure}[!b]
 \begin{center}
 \includegraphics[scale=0.23]{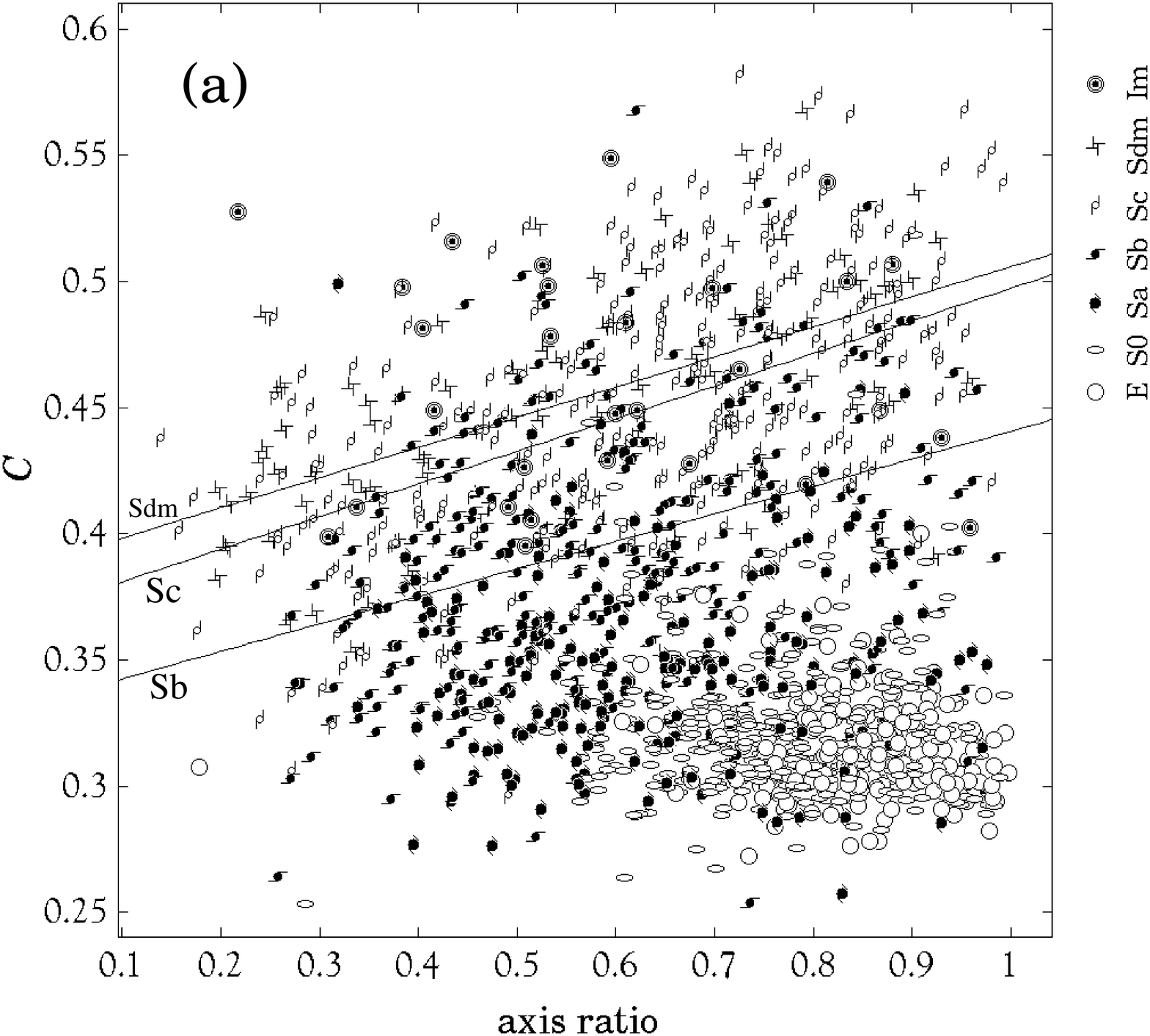} \\
 \includegraphics[scale=0.23]{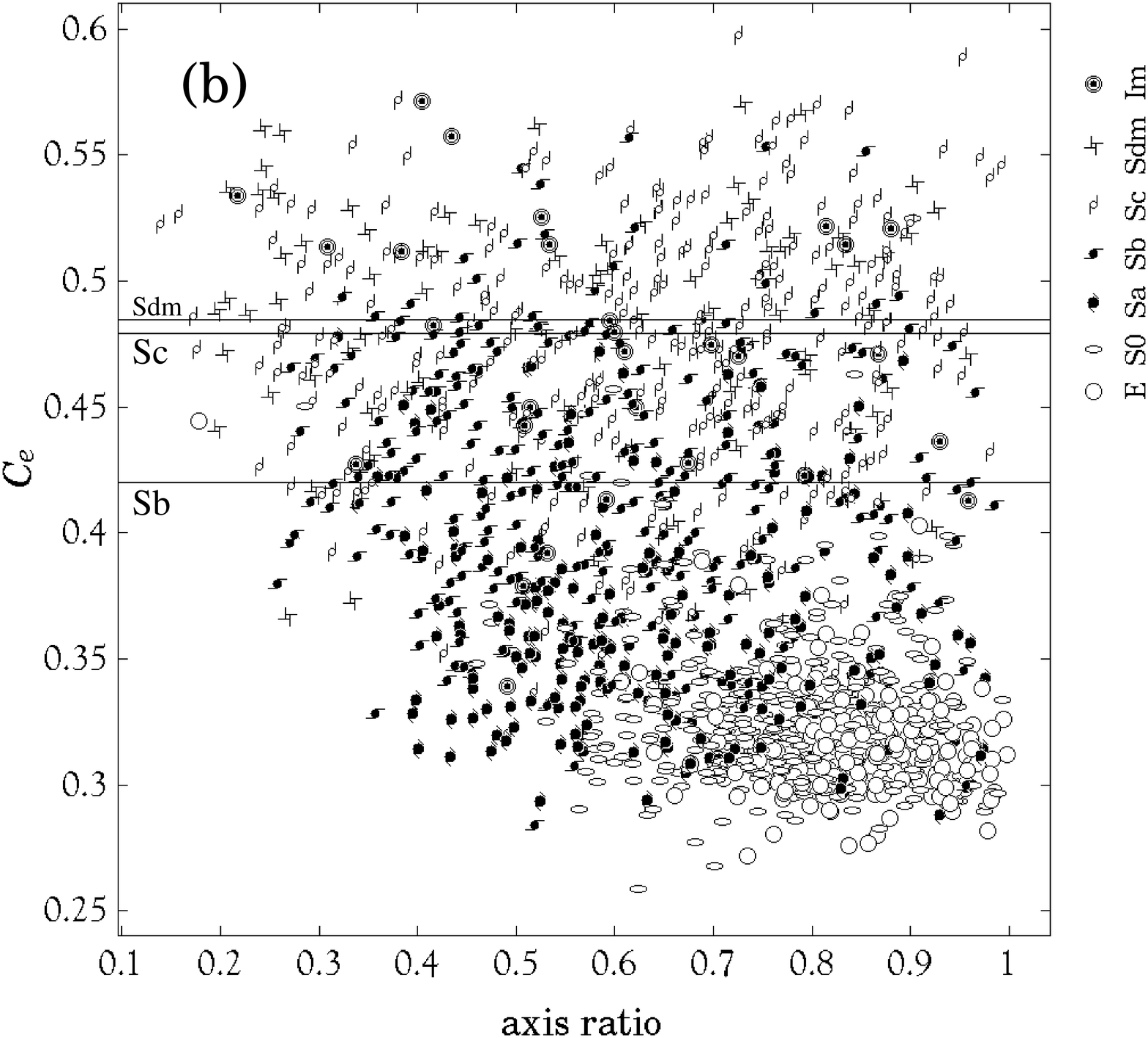}
 \vspace*{-9mm}
 \end{center}
 \caption{Concentration index versus axis ratio using circular
(a) and elliptical (b) aperture definitions. 
%Inverse concentration indices plotted against 
% axis ratio $\alpha$.
 Each symbol indicates visual classification $T$ as specified in
the legend, and
 the straight lines are the linear regression lines for types Sb, Sc and Sdm.
% (a) Figure with circular apertures;
%  (b) with elliptical apertures.
}
 \label{alpha_vs_cin}
\end{figure}

%We defined the elliptical concentration index $C_e$ 
%in section \ref{Ce} 
%to recover the disadvantage of $C$.
We show that smaller axis ratio affects 
the standard concentration index $C=r_{50}/r_{90}$ 
defined with the Petrosian flux in the circular
apertures,
and using the elliptical ones
can remove the effect significantly.

In Figure \ref{t_vs_cin} we plot
the two concentration indices
against visual morphological type index $T$, 
%employing
calculated by
(a) conventional circular apertures ($C$), and 
(b) elliptical apertures ($C_e$). 
Galaxies that follow de Vaucouleurs' law give $C$=$C_e$=$0.29$ and those
with the exponential profile give 0.44.

The Spearman correlation coefficient with the use of $C$, $r$=$0.781$
%,
increases
to $r$=$0.820$ with $C_e$. 
%Several significant improvements are obtained by the use of an elliptical 
%aperture ($C_e$), compared with the original circular aperture ($C$):
%For example, 
%those galaxies with 3$\le$$T$$\le$3.5 and 0.3$\le$$C$$\le$0.45
%in panel (a) (under the dotted line) 
%have much larger value of $C_e$ in panel (b). These are edge-on 
%galaxies
%mis-classified
%using the circular aperture,
%but correctly classified as less concentrated galaxies
%using the elliptical aperture. 
Several significant improvements are obtained by the use of an
elliptical aperture ($C_e$) compared with the original circular aperture
($C$). From Figure \ref{t_vs_cin}, the number of spiral galaxies
visually classified as $T$=$3$ and $T$=$3.5$ under the dotted line
is smaller elliptical than circular
concentration index estimates. This is expected for inclined
later galaxy types since according to our definition less concentrated
galaxies have larger concentration indices. The sample which
exhibits larger difference between panel (a) and (b) in Figure \ref{t_vs_cin}
contains a significant number of edge-on spirals, which are
misclassified using the
circular aperture, but correctly identified as less concentrated using
the elliptical aperture estimates.
A similar improvement is also seen for Sa galaxies ($T$=2).

Thus, using elliptical aperture concentration indices for
morphological classification increases the classification accuracy by
making it independent of galaxy inclination. In Figure
\ref{alpha_vs_cin}
we present
the concentration index versus the axis ratio (smaller axis ratios are
indicative of larger inclinations) for both the circular (panel (a))
and elliptical aperture (panel (b)) concentration indices. 
%Figure \ref{alpha_vs_cin} presents
%the concentration index against the axis ratio. 
The regression lines are drawn for morphological
classes of galaxies Sb, Sc and Sdm.
The lines obtained by regression on
samples of different galaxy types %shown 
in panel (a), which are based on $C$, 
are significantly tilted  with respect to the axis ratio.
%The concentration indices of late-type galaxies, when defined with
%circular apertures, drop to the values of earlier types
%when they are largely inclined. 
The concentration indices of highly-inclined late-type galaxies
estimated using circular apertures are artificially reduced to values
characteristic of elliptical morphologies.
That is, edge-on spirals give an anomalously high concentration of light
if defined with circular apertures; the 
morphological classifications
that use only $C$ have significant contaminants.
% than they really appear
%in the image.
%This means that
%the morphological classification with the use of only
%$C$ has a significant contaminants.
%
Figure \ref{alpha_vs_cin}(b), using $C_e$, shows that elliptical
apertures remedy
this problem.  We observe
little tilt of the regression lines, suggesting that
$C_e$ works better to classify morphologies than $C$;
the effect of inclination 
is removed with the use of elliptical apertures.

\subsection{Coarseness parameter} 
\label{behavior_of_Y}

\begin{figure}[!b]
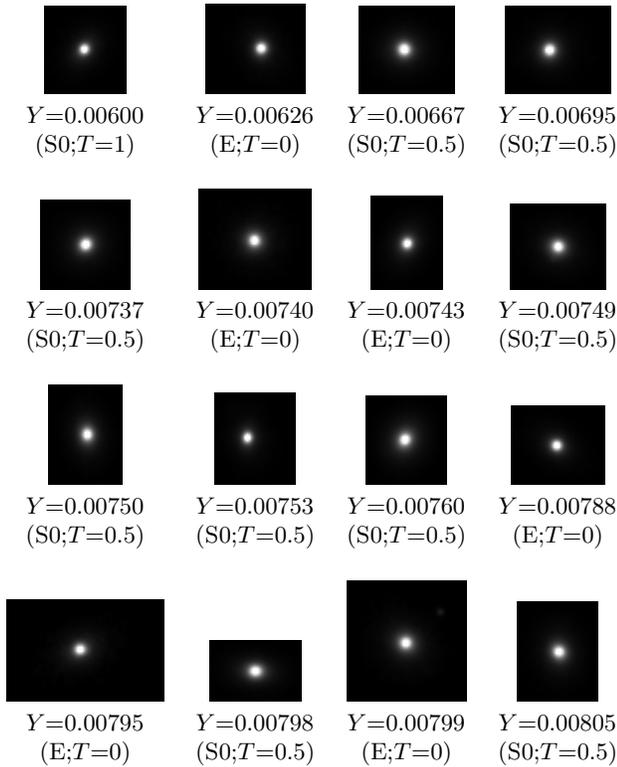

 \begin{center}
{\small
 \begin{tabular}{cccc}
  \includegraphics[scale=0.8]{ChisatoYamauchi.fig8a.eps3} &
  \includegraphics[scale=0.8]{ChisatoYamauchi.fig8b.eps3} &
  \includegraphics[scale=0.8]{ChisatoYamauchi.fig8c.eps3} &
  \includegraphics[scale=0.8]{ChisatoYamauchi.fig8d.eps3} \\
  $Y$=0.00600 & $Y$=0.00626 & $Y$=0.00667 & $Y$=0.00695 \\
  (S0;$T$=1) & (E;$T$=0) & (S0;$T$=0.5) & (S0;$T$=0.5) \\
    & & & \\
  \includegraphics[scale=0.8]{ChisatoYamauchi.fig8e.eps3} &
  \includegraphics[scale=0.8]{ChisatoYamauchi.fig8f.eps3} &
  \includegraphics[scale=0.8]{ChisatoYamauchi.fig8g.eps3} &
  \includegraphics[scale=0.8]{ChisatoYamauchi.fig8h.eps3} \\
  $Y$=0.00737 & $Y$=0.00740 & $Y$=0.00743 & $Y$=0.00749 \\
  (S0;$T$=0.5) & (E;$T$=0) & (E;$T$=0) & (S0;$T$=0.5) \\
    & & & \\
  \includegraphics[scale=0.8]{ChisatoYamauchi.fig8i.eps3} &
  \includegraphics[scale=0.8]{ChisatoYamauchi.fig8j.eps3} &
  \includegraphics[scale=0.8]{ChisatoYamauchi.fig8k.eps3} &
  \includegraphics[scale=0.8]{ChisatoYamauchi.fig8l.eps3} \\
  $Y$=0.00750 & $Y$=0.00753 & $Y$=0.00760 & $Y$=0.00788 \\
  (S0;$T$=0.5) & (S0;$T$=0.5) & (S0;$T$=0.5) & (E;$T$=0) \\
    & & & \\
  \includegraphics[scale=0.8]{ChisatoYamauchi.fig8m.eps3} &
  \includegraphics[scale=0.8]{ChisatoYamauchi.fig8n.eps3} &
  \includegraphics[scale=0.8]{ChisatoYamauchi.fig8o.eps3} &
  \includegraphics[scale=0.8]{ChisatoYamauchi.fig8p.eps3} \\
  $Y$=0.00795 & $Y$=0.00798 & $Y$=0.00799 & $Y$=0.00805 \\
  (E;$T$=0) & (S0;$T$=0.5) & (E;$T$=0) & (S0;$T$=0.5) \\
 \end{tabular}
}
 \end{center}
 \caption{%The $g$-band images (rescaled) of 16 galaxies having the lowest
% $Y$ in our 1421 galaxies. Shown in parentheses are the visually 
%classified morphological types.
Rescaled $g$-band images of the 16 galaxies with the lowest coarseness
parameters in our sample ($Y \leq 0.00805$). The visually classified
morphological types are shown in brackets together with the quantitative
morphological index $T$.
}
 \label{img_earytypes}
\end{figure}

%\clearpage

\begin{figure}[!b]
 \begin{center}
{\small
 \begin{tabular}{cccc}
  \includegraphics[scale=0.8]{ChisatoYamauchi.fig9a.eps3} &
  \includegraphics[scale=0.8]{ChisatoYamauchi.fig9b.eps3} &
  \includegraphics[scale=0.8]{ChisatoYamauchi.fig9c.eps3} &
  \includegraphics[scale=0.8]{ChisatoYamauchi.fig9d.eps3} \\
  $Y$=0.378 & $Y$=0.336 & $Y$=0.330 & $Y$=0.321 \\
  (Sc;$T$=4) & (Sc;$T$=4) & (Im;$T$=6) & (Sdm;$T$=5) \\
    & & & \\
  \includegraphics[scale=0.8]{ChisatoYamauchi.fig9e.eps3} &
  \includegraphics[scale=0.8]{ChisatoYamauchi.fig9f.eps3} &
  \includegraphics[scale=0.8]{ChisatoYamauchi.fig9g.eps3} &
  \includegraphics[scale=0.8]{ChisatoYamauchi.fig9h.eps3} \\
  $Y$=0.321 & $Y$=0.304 & $Y$=0.303 & $Y$=0.293 \\
  (Sc;$T$=3.5) & (Sc;$T$=3.5) & (Sc;$T$=4) & (Sc;$T$=4) \\
    & & & \\
  \includegraphics[scale=0.8]{ChisatoYamauchi.fig9i.eps3} &
  \includegraphics[scale=0.8]{ChisatoYamauchi.fig9j.eps3} &
  \includegraphics[scale=0.8]{ChisatoYamauchi.fig9k.eps3} &
  \includegraphics[scale=0.8]{ChisatoYamauchi.fig9l.eps3} \\
  $Y$=0.290 & $Y$=0.275 & $Y$=0.273 & $Y$=0.269 \\
  (Sdm;$T$=4.5) & (Sc;$T$=4) & (Sc;$T$=4) & (Sc;$T$=4) \\
    & & & \\
  \includegraphics[scale=0.8]{ChisatoYamauchi.fig9m.eps3} &
  \includegraphics[scale=0.8]{ChisatoYamauchi.fig9n.eps3} &
  \includegraphics[scale=0.8]{ChisatoYamauchi.fig9o.eps3} &
  \includegraphics[scale=0.8]{ChisatoYamauchi.fig9p.eps3} \\
  $Y$=0.268 & $Y$=0.266 & $Y$=0.264 & $Y$=0.260 \\
  (Sc;$T$=3.5) & (Sc;$T$=3.5) & (Sc;$T$=3.5) & (Im;$T$=6) \\
 \end{tabular}
}
 \end{center}
 \caption{
%The $g$-band images (rescaled) of 16 galaxies having the largest
% $Y$ in our 1421 galaxies. Shown in parentheses are the visually 
%classified morphological types.
Rescaled $g$-band images of the 16 galaxies with the largest coarseness
parameters in our sample ($0.260 \leq Y$). The visually classified
morphological types are shown in brackets together with the quantitative
morphological index $T$.
}
 \label{img_latetypes}
\end{figure}

%\clearpage

\begin{figure}[!t]
 \begin{center}
 \includegraphics[scale=0.25]{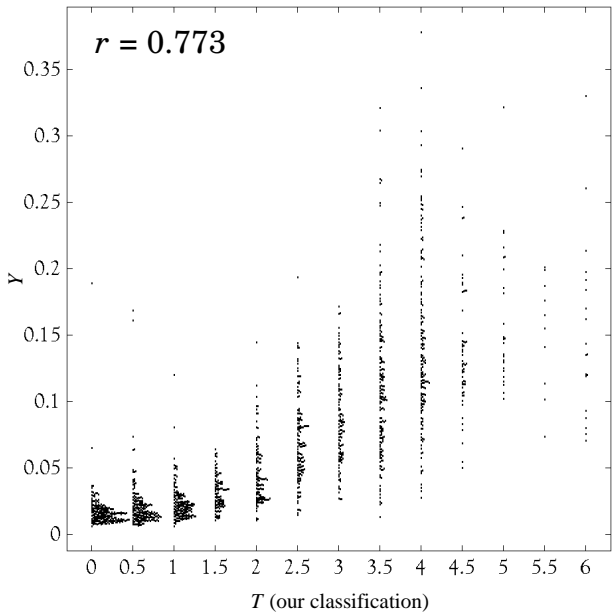}
 \end{center}
 \caption{
%The coarseness parameter $Y$ of 1421 galaxies 
% plotted against morphology $T$.
% The value at the top of the figure is the linear correlation coefficient.
%In case more than one galaxy has the same ($T$,$X$) value, $T$ values are
%displaced to the right by a dot size for clarity. In other words, this
%scatter plot can be viewed as a histogram.
The coarseness parameter $Y$ 
of the 1421 galaxies in our sample versus the visual morphology
index $T$.
Some points were shifted to larger $T$ values to avoid overlaps. The
linear correlation
coefficient is given in the top left corner of the panel. 
}
 \label{t_vs_cn}
\end{figure}

\begin{figure}[!b]
 \begin{center}
 \includegraphics[scale=0.234]{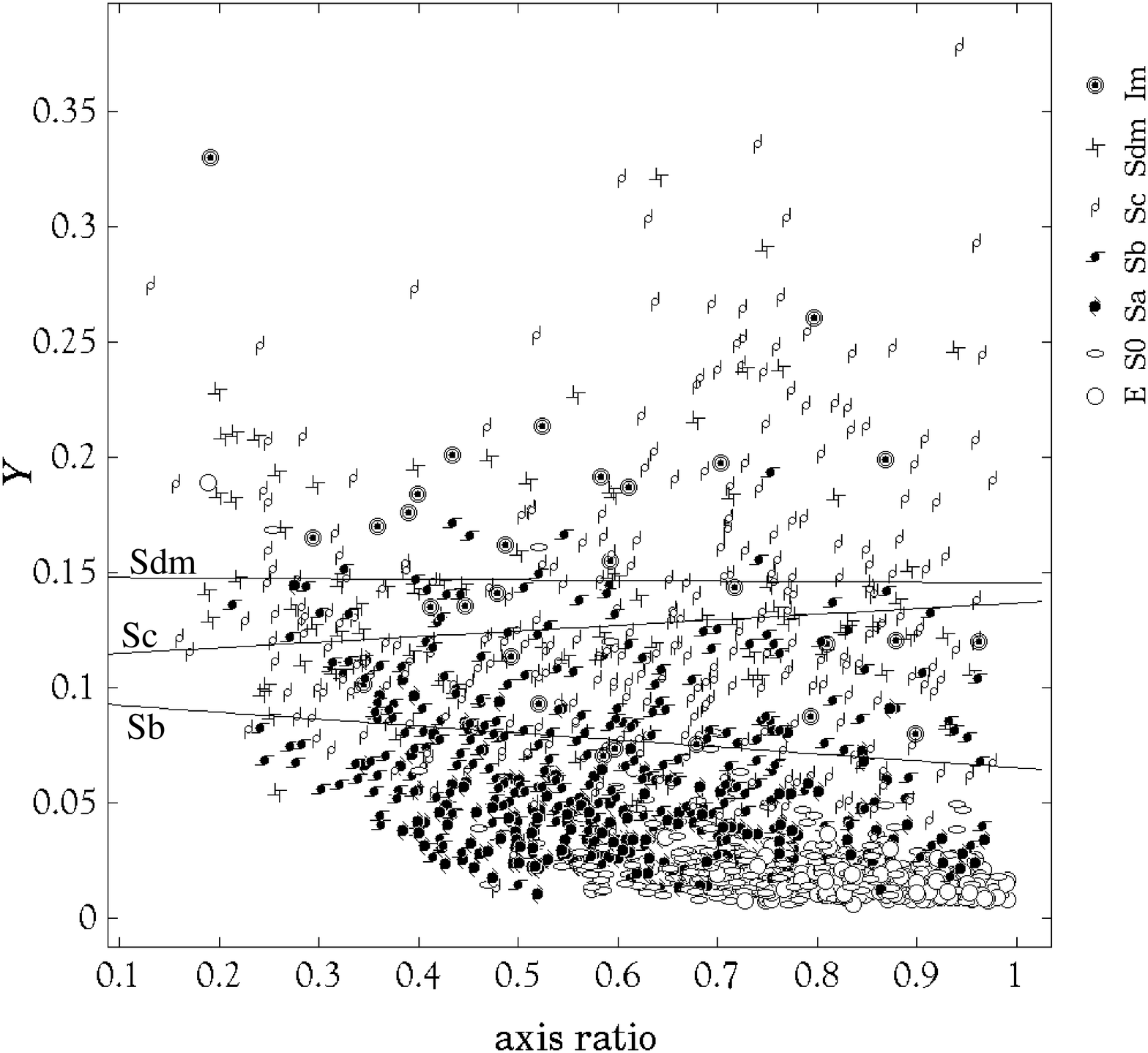}
 \end{center}
 \caption{The coarseness parameter $Y$ plotted against the axis ratio.
 Each symbol indicates visual morphological type, and
 straight lines are the linear regression for Sb, Sc and Sdm galaxies.}
 \label{alpha_vs_cn}
\end{figure}

%\clearpage

%The coarseness parameter $Y$ is defined as the
%equation (\ref{eq_of_Y}) which exhibits the
%ratio of the range of fluctuations in surface brightness 
%along the elliptic circumference to the full dynamic range of surface
%brightness of the galaxy. 
%We show what galaxy image makes $Y$ parameter larger or smaller, 
%and the correlation between $Y$ parameter and the visual morphological type
%$T$.
%
%We display 16 rescaled $g$-band images of galaxies which have 
%low $Y$ values ($Y$$<$$0.00805$) in Figure \ref{img_earytypes}, 
%and another 16 having
%high $Y$ values ($Y$$>$$0.260$) in Figure \ref{img_latetypes}. 
The coarseness parameter $Y$, as defined in equation (\ref{eq_of_Y}), is
equal to
the ratio of the range of fluctuations in surface brightness
(along an elliptical circumference) to the full dynamic range of
surface brightness. Larger values of $Y$ are indicative of the presence
of structure in the galaxy disks (e.g., spiral arms) and are associated
with galaxies visually classified as late types.  Figures \ref{img_earytypes} and \ref{img_latetypes} each
display 16 galaxies with the smallest ($Y$$<$$0.00805$) and the largest
($Y$$>$$0.260$) $Y$ parameters present in our sample.
It is apparent
that the galaxies shown in Figure \ref{img_earytypes} possess very weak
texture and are classified as early types. 
All galaxies shown in Figure \ref{img_latetypes} have conspicuous
texture and are indeed classified as late (Sbc or later)-type spirals.

It is obvious that texture of galaxies contributes to the numerator
of equation (\ref{eq_of_Y}). We should, however, emphasize the role of the denominator. 
When galaxies have conspicuous bulges, their $\Delta I$ values
suppress the $Y$ parameter. On the other hand, the faint bulge leads to
a small $\Delta I$ value that enhances $Y$.
%The most typical case is that of
%Magellanic irregular type galaxies, in which texture is not very 
%conspicuous, but the overall intensity contrast is small, giving a small 
%denominator so that $Y$ indicates a very late type. 
A typical example of the importance of the denominator of equation
(\ref{eq_of_Y}) is a type of galaxy known as a Magellanic irregular. The
texture of Magellanic irregulars is not conspicuous, but the overall
intensity contrast is also small, resulting in a small denominator in
equation
(\ref{eq_of_Y}) and large $Y$ parameter indicative of a very late type.

Figure \ref{t_vs_cn} displays the coarseness parameter $Y$
plotted against $T$. The correlation ($r$=$0.773$) is not very
impressive compared to 
that for $C_e$-$T$, %but this %somewhat small 
%$r$ is caused by
%the fact that 
since
the $Y$-$T$ correlation is curved away from
the linear relation. The important feature in this figure is a very narrow
distribution of the $Y$ parameter for 0$\le$$T$$\le$1 galaxies.
The distribution is confined to the range 0$\le$$Y$$\la$0.03, 
which is but 10\% of the full variation of $Y$.
%1/10 the full range of $Y$ from early- to late-type galaxies.

The %effect of inclination on the $Y$ 
relation between the $Y$ and axis ratio
is shown in 
Figure \ref{alpha_vs_cn}. 
The linear regression lines for different $T$ are almost flat.
They do not appear as flat as those
for $C_e$, but 
%the tilt is not systematic and 
%seems to be caused by accidental effects.  
the slopes of the lines are not caused by any systematic
effects, i.e., Sdm is almost flat and Sc/Sb have opposite slopes.
%but by the random scatter of the data.
%We conclude that $Y$ does not receive the systematic effect of inclination.
We conclude that $Y$ is not affected severely by the inclination of galaxies.

\section{MORPHOLOGICAL CLASSIFICATIONS USING $C_{(e)}$ and $Y$ INDICES}
\label{classification}

%We now study the performance of classification using different indicators
%taking visual classification as a reference.

\subsection{Early versus late types}
\label{results}

\begin{figure}[!b]
 \begin{center}
 \includegraphics[scale=0.56]{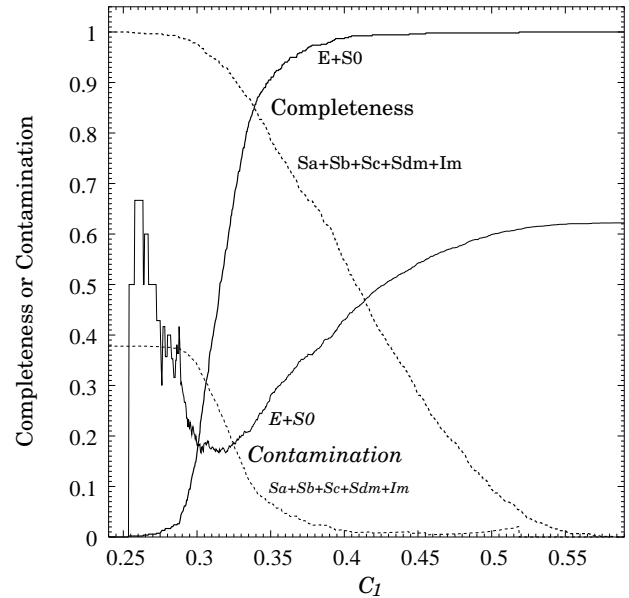}
 \vspace*{-3mm}
 \end{center}
 \caption{Completeness $P$ and contamination $Q$ for the early- and
  late-type galaxy samples as a function of the division parameter $C_{1}$
   with the standard circular aperture definition. 
  The thick solid and dotted lines show completeness for
  early- and late-types.
  The thin solid and dotted lines show contamination for
  early- and late-types.
 }
 \label{com_cin_E}
\end{figure}

We first attempt to classify galaxies into two types, early %`$e$'
(E-S0/a) and
late %`$\ell$'
(Sa-Im),
%The galaxies are classified into two classes 
by setting a dividing value for each of three parameters 
displayed in Figure \ref{t_vs_cin}(a,b) and Figure \ref{t_vs_cn}.
We call the two classes of the sample 
`$e$' and `$\ell$'.
We evaluate the completeness $P$ and the contamination $Q$ of the `$e$'
and `$\ell$', as defined by
\begin{equation}
 P_{e} = \frac{N_{e\{{\rm E+S0\}} }} { N_{\rm \{E+S0\}} }~,~~~
 Q_{e} = \frac{N_{e {\rm\{Sa+Sb+Sc+Sdm+Im\}}}} { N_{e }}~,
\end{equation}
\begin{equation}
 P_{\ell} = \frac{N_{\ell {\rm\{Sa+Sb+Sc+Sdm+Im\}}}} { N_{\rm \{Sa+Sb+Sc+Sdm+Im\}} }~,~~~
 Q_{\ell} = \frac{N_{\ell{\rm \{E+S0\}}}} { N_{\ell }}~,
\end{equation}
%where $N_{e {\{{\rm E+S0}\}} }$ is the number of E+S0 galaxies in 
%the `$e$' sample, and $N_{\{{\rm E+S0}\}}$ is
%the total number of E+S0 galaxies. 
where $N_{e}$ is the number of all galaxies chosen by the separator line,
$N_{e {\{{\rm E+S0}\}} }$ is the number of E+S0 galaxies 
chosen by the separator line, and $N_{\{{\rm E+S0}\}}$ is
the total number of E+S0 galaxies. 
Other notations are defined similarly.

\begin{figure}[!b]
 \begin{center}
 \includegraphics[scale=0.56]{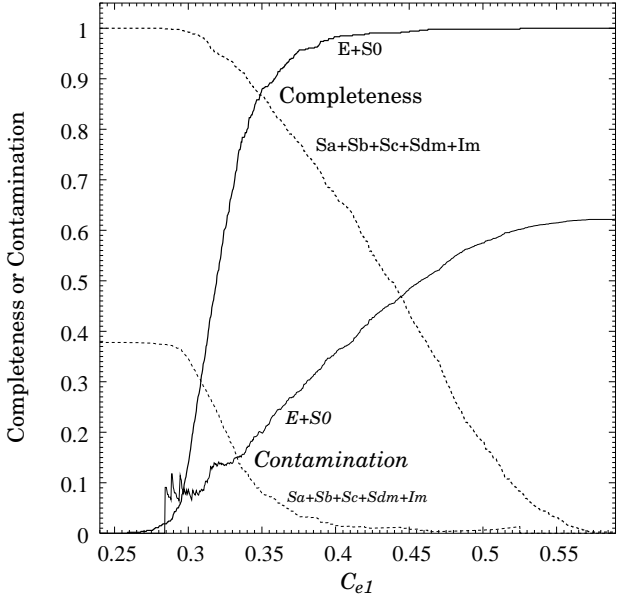}
 \vspace*{-4mm}
 \end{center}
 \caption{Completeness $P$ and contamination $Q$ of the early- and
  late-type galaxy samples as a function of the division parameter $C_{e1}$,
  defined using elliptical apertures.
  See Figure \ref{com_cin_E} for the line definitions.
 }
 \vspace*{4mm}
 \label{com_ci_E}
\end{figure}

\begin{figure}[!b]
 \begin{center}
 \includegraphics[scale=0.56]{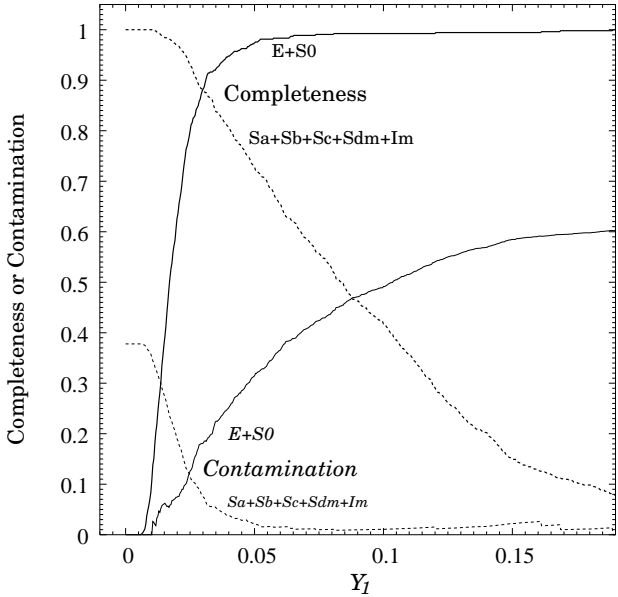}
 \vspace*{-4mm}
 \end{center}
 \caption{Completeness $P$ and contamination $Q$ of the early- and
  late-type galaxy samples as a function of the division parameter $Y_{1}$.
  See Figure \ref{com_cin_E} for the line definitions.
 }
 \label{com_ci+cn_E}
\end{figure}

\begin{figure}[!b]
 \begin{center}
 \includegraphics[scale=0.56]{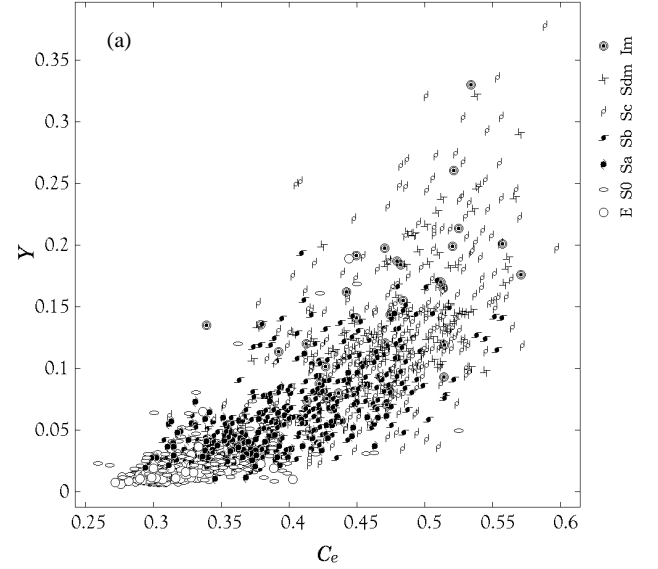}
 \includegraphics[scale=0.56]{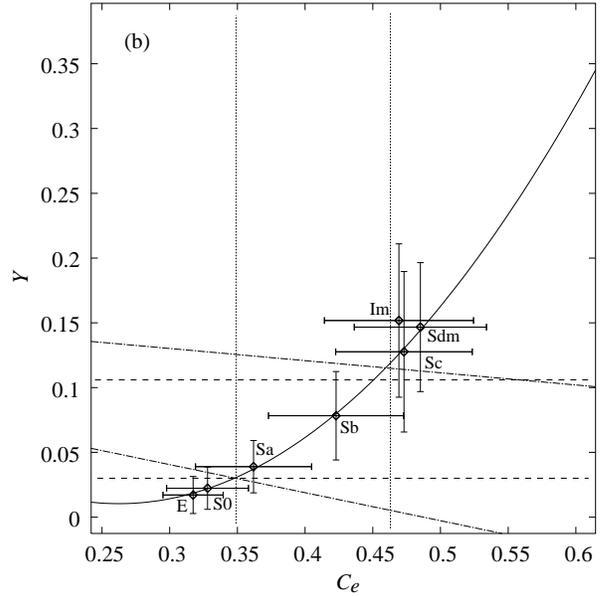}
 \vspace*{-8mm}
 \end{center}
 \caption{(a) 1421 Galaxies plotted on the ($C_e$,$Y$) plane and
 (b) the average 2-vector and standard deviation for each morphological
 class.
%The quadratic function in (b) is the regression line by
%least-squares fitting to the average 2-vector of galaxies.
The solid line in
panel (b) is the quadratic function least-squares fit to the points
representing the average values for each morphological class.
The two dot-dashed lines are the best classifiers into
three types, E+S0, Sa+Sb and Sc+Sdm+Im using the 2-dimensional
classification.
The dotted and dashed lines 
%show the best 
%of one dimensional separators using $C_e$ and $Y$, respectively.
show the best one dimensional separators using
$C_e$ and $Y$, respectively.
}
 \label{ci_vs_cn}
\end{figure}

Figure \ref{com_cin_E} shows completeness and contamination for
the classification using the $C$ parameter. 
The same analysis is performed in
\citet{shi01}
which attains 80\% completeness at $C_1$=$0.35$.
%but based on %a different galaxy sample from
%the SDSS commissioning data.
\citet{str01} also reports 83\% completeness at $C_1$=$0.38$
using a concentration index with circular apertures.
We find that 
$P_{e}$=$P_{\ell}$=$85.0$\%, $Q_e$=$22.5$\% and $Q_{\ell}$=$9.5$\%
with the use of the division constant
$C_1$=$0.339$. The success rate is somewhat higher in our case, however,
with essentially the same division parameter.
Improved performance is seen in Figure \ref{com_ci_E}
where the $C$ parameter is replaced with $C_e$: we obtain
$P_{e}$=$P_{\ell}$=$86.7$\%, $Q_e$=$20.0$\% and $Q_{\ell}$=$8.2$\%
for $C_{e1}$=$0.349$. 
%The use of ellipse-based concentration index 
%improves morphological classifications.
One may attain $Q_e$=$Q_\ell$=$15.0\%$ with the choice of $C_{e1}$=$0.332$
if the completeness of E+S0 galaxies is sacrificed.

\begin{figure}[!b]
 \begin{center}
 \includegraphics[scale=0.56]{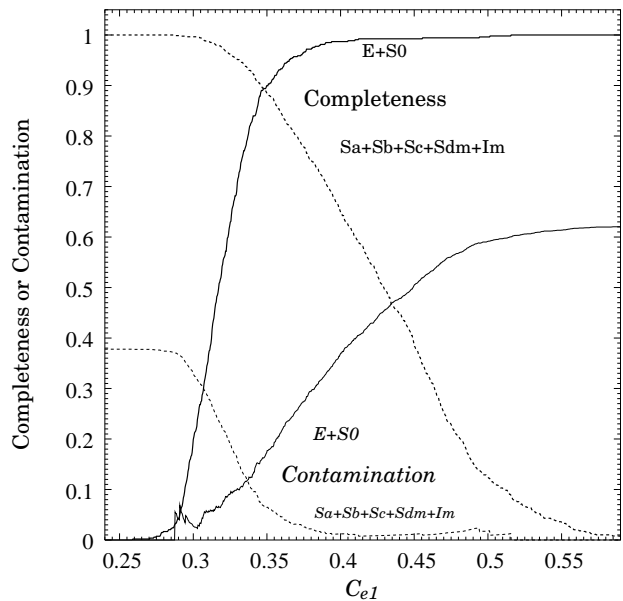}
 \vspace*{-5mm}
 \end{center}
 \caption{Completeness and contamination of the early and
  late-type galaxy samples as a function of $C_{e1}$ specified 
  at the crossing point
  with the quadratic function,  
 using the $C_e$-$Y$ diagram (Figure \ref{ci_vs_cn}).
  See Figure \ref{com_cin_E} for the line definitions.
 }
 \label{com_ce+y_E}
\end{figure}

The result of classification using the $Y$ parameter is presented in 
Figure \ref{com_ci+cn_E}. This result shows a higher success rate
than the $C_e$ classification,
$P_{e}$=$P_{\ell}$=$ 88.1\%$, $Q_{e}$=$18.1$\% and $Q_{\ell}$=$7.5\%$
for $Y_1$=$0.030$.  We can attain minimum contamination of
$Q_e$=$Q_\ell$=$12.1\%$ for $Y_1$=$0.024$ with a modest cost of $P_e$.
%The $Y$ parameter is superior to 
We conclude that the $Y$ parameter is superior to 
the concentration indices
as the morphology classifier.

We now try to obtain the maximum performance using two parameters,
$C_e$ and $Y$, by optimising the choice of the 
dividing parameters (see Figure \ref{ci_vs_cn}).  
We consider the position of the center
of the average 2-vector for each morphological class,
\begin{eqnarray}
%\footnotesize
 {G}_{\rm E} = 
 \left( \matrix{ \langle {{C_e}_{\rm E}} \rangle \cr 
  \langle {Y_{\rm E} \rangle } } \right),~
 {G}_{\rm S0} = 
 \left( \matrix{ \langle {{C_e}_{\rm S0}} \rangle \cr \langle
 {Y_{\rm S0} \rangle} }
 \right), \nonumber\\
 ... ,~ 
 {G}_{\rm Im} = 
 \left( \matrix{ \langle {{C_e}_{\rm Im}} \rangle \cr \langle
 {Y_{\rm Im} \rangle} } \right) .
\end{eqnarray}
where $\langle {{C_e}_{\rm E}}\rangle=
(1/N_{\rm E})\sum_{i\in {\rm E}}{{C_e}_{\rm E}^i}$
etc. 
With the weight of the number of the galaxies having the relevant 
morphological class given to these points,
we fit 7 points by a quadratic function. For our 1421 galaxies,
we obtain 
\begin{equation}
 f_q(C_e) = 2.702 C_e^2 -1.419 C_e +0.1967 .
\end{equation}

%In general, when the distribution in two-parameter space shows 
%linear regression, the parallel separator lines are placed along 
%a regression line (Doi et al. 1993).  
In general, for 2-D distributions that show separate linear
correlations for each type, the separator lines are parallel to the
individual correlations (see Doi et al. 1993).
We apply this method to
the quadratic regression line.
We then consider a set of lines crossing this quadratic curve at
$(C_{e1},f_q(C_{e1}))$,
\begin{equation}
f_l(C_e) = - \frac{K}{f_q'(C_{e1})}(C_e - C_{e1}) + f_q(C_{e1})
\label{eq:CYdivline}
\end{equation}
where $K$ is a constant, and
$f_q'$ is the first derivative of $f_q$.
We take $f_l(C_e)$ as the dividing line for 
classification. We 
adopt $K$=$0.1$, which turns out to give the best performance.
%Figure \ref{ci_vs_cn} also shows $f_l(C_{e1})$, among which the
%lines we adopt as dividers are denoted by dotted symbols.
Figure \ref{ci_vs_cn} shows
%average 2-vector of galaxies classified into 7 morphological types,
the average 2-vector and standard deviation for each morphological class,
%quadratic regression $f_q(C_e)$ and the best classifiers by 
%equation (\ref{eq:CYdivline}) plotted as two dot-dashed separator lines.
the quadratic regression $f_q(C_e)$ and the best separator line
between the early and late type galaxies (lower dot-dashed line).

%The result of classification using the $C_e$-$Y$ diagram is presented in 
%Figure \ref{com_ce+y_E}. 
The completeness and contamination curves for classification using
the $C_e$-$Y$ diagram are presented in Figure \ref{com_ce+y_E}.
%The maximum success rate we achieved is:
%$P_{e}$=$P_{\ell}$=$89.4\%$, while $Q_{e}$=$16.4\%$ and $Q_{\ell}$=$ 6.7\%$
%at $C_{e1}({\rm crossing})$=$0.348$, or $Q_{e}$=$Q_{\ell}$=$11.8\%$ with
%$P_{e}$=$79.3$\% and $P_{\ell}$=$93.6$\% at $C_{e1}({\rm crossing})$=$0.337$.
The maximum success rate we achieved is (assuming equal completeness
for both early and late types) $P_{e}$=$P_{\ell}$=$89.4\%$, 
with $Q_{e}$=$16.4\%$ and $Q_{\ell}$=$ 6.7\%$
for $C_{e1}$=$0.348$, or (assuming equal contamination)
$P_{e}$=$79.3$\% and $P_{\ell}$=$93.6$\%
with $Q_{e}$=$Q_{\ell}$=$11.8\%$ for $C_{e1}$=$0.337$.
These are remarkably high success rates given the fact that
visual classification suffers from uncertainties, perhaps of the order
$\Delta T\sim 1$.

\subsection{Classification into three types}

\begin{table}[!b]
\begin{center}
 \begin{tabular}{ccccc}
\hline
\hline
 \multicolumn{5}{c}{$C$ Parameter} \\
\hline
                         & E+S0 & Sa+Sb & Sc+Sdm+Im & Total \\
  $e$             & {\bf 458 } & 124        &  9        & 591 \\
  $\ell_{\rm I}$  & 76         & {\bf 278 } & 161       & 515\\
  $\ell_{\rm II}$ &  3         & 59         & {\bf 253 }& 315\\
  Total                  & 537        & 461        & 423       & 1421\\
\hline
\hline
 \multicolumn{5}{c}{$C_e$ Parameter} \\
\hline
                         & E+S0 & Sa+Sb & Sc+Sdm+Im & Total \\
  $e$             & {\bf 468 } & 111         &  6        & 585 \\
  $\ell_{\rm I}$  & 68         & {\bf 285 }  & 156       & 509 \\
  $\ell_{\rm II}$ &  1         & 65          & {\bf 261 }& 327 \\
  Total                  & 537        & 461         & 423       & 1421\\
\hline
\hline
 \multicolumn{5}{c}{$Y$ Parameter} \\
\hline
                         & E+S0 & Sa+Sb & Sc+Sdm+Im & Total \\
  $e$             & {\bf 474 } & 99         &  6        & 579 \\
  $\ell_{\rm I}$  & 59         & {\bf 305 } & 138       & 502 \\
  $\ell_{\rm II}$ &  4         & 57         & {\bf 279 }& 340 \\
  Total                  & 537        & 461        & 423       & 1421\\
\hline
 \end{tabular}\\
\end{center}
 \caption{%
Classification of 1421 galaxies 
into three types (E+S0 : Sa+Sb : Sc+Sdm+Im)
using parameter $C$,$C_e$, or $Y$.
%A completeness can be calculated by a column,
%and a contamination by a row; i.e. in the case of $C$,
The completeness of a
selection method can be calculated using the numbers from each column
as follows. 
%For the selection method using $C$,
%completeness of E+S0 in the $e$
%is 458/537$\simeq$0.85 (by a column), and contamination
%of the $e$ is (124+9)/591$\simeq$0.23 (by a row).
%
For the selection method using $C$, the completeness of the E+S0 sample
is 458/537$\simeq$0.85 (from column 1 of the $C$ table), and the contamination
is (124+9)/591$\simeq$0.23 (from row 1 of the $C$ table).
}
 \label{3x3tbl_1}
\end{table}

\begin{table}[!b]
\begin{center}
 \begin{tabular}{ccccc}
\hline
\hline
 \multicolumn{5}{c}{$C_e$-$Y$ Diagram}     \\
\hline
                         & E+S0 & Sa+Sb & Sc+Sdm+Im & Total \\
  $e$             & {\bf 480 } & 92         &  2        & 574 \\
  $\ell_{\rm I}$  & 53         & {\bf 314 } & 137       & 504 \\
  $\ell_{\rm II}$ &  4         & 55         & {\bf 284 }& 343 \\
  Total                  & 537        & 461        & 423       & 1421\\
\hline
 \end{tabular}\\
\end{center}
 \caption{%
Classification of 1421 galaxies 
into three types (E+S0/Sa+Sb/Sc+Sdm+Im)
using the $C_e$-$Y$ diagram.
%A completeness can be calculated by a column,
%and a contamination by a row; i.e. 
The completeness of a
selection method can be calculated using the numbers from each 
column as follows. The
completeness of E+S0 sample in the $e$
is 480/537$\simeq$0.89 (from column 1), and the contamination
of the $e$ is (92+2)/574$\simeq$0.16 (from row 1).}
 \label{3x3tbl_2}
\end{table}

We consider separation into three types, dividing late (`$\ell$')- 
type galaxies into early ($\sim$ Sa+Sb) and late spirals
($\sim$ Sc+Sdm+Im). 
We call the two classes of the sample 
$\ell_{\rm I}$ and $\ell_{\rm II}$.
We fix the first division 
%to give the best performance for
for the early/late
%separation
classification
as determined in the previous subsection.
We set the second division which separates 
$\ell_{\rm I}$ and $\ell_{\rm II}$
so that the completeness of Sa+Sb in $\ell_{\rm I}$ and 
that of Sc+Sdm+Im in $\ell_{\rm II}$
are nearly equal.

We show the results in matrices of $3 \times 3$
(and an additional row and column to show the subtotals)
in Table \ref{3x3tbl_1} for the three indicators 
using $C$, $C_e$ and $Y$. 
We take the second division separators which divide late-type galaxies
into early- and late-spirals
to be $C_{2}$=$0.436$, $C_{e2}$=$0.463$, and $Y_2$=$0.106$, respectively,
while the separators between early- and late-type
galaxies are the same as those quoted in the beginning of Section \ref{results}.
The completeness can be read from the column by dividing the
number in the diagonal entry by the total number of listed 
in the bottom of the corresponding column.
For example, the completeness of Sa+Sb galaxies
separated by $C_{1,2}$ is $278/461=60.3$\%.
This compares to $285/461=61.8$\% with the use of $C_e$,
and $305/461=66.2$\% with $Y$.
The contamination is read from the row.
The contamination in the early spiral galaxy sample, 
$\ell_{\rm I}$, from E+S0 galaxies and late-type spiral galaxies,
for instance, is $(76+161)/515=46.0$\% with the use of $C$,
$(68+156)/509=44.0$\% with $C_e$, and 
$(59+138)/502=39.2$\% with $Y$.
%Similarly, the 
%contamination in the late spiral $\ell_{\rm II}$ sample
%is $20.0$\%, $20.2$\%, and $17.9$\% respectively.
Similarly the $\ell_{\rm II}$ %sample
contamination is $20.0$\%, $20.2$\%, and $17.9$\%, for the $C$, $C_e$, and
$Y$ parameter classification, respectively.

The classification with $Y$ again produces the best result. 
We also note that the gain attained with the
use of elliptical apertures over circular apertures is small 
for %the separation of 
$\ell_{\rm I}$ and $\ell_{\rm II}$, 
although the $C_e$ parameter
gives generally better performance, if slight. One might question
why the gain with $C_e$ over $C$ is rather small 
in contrast to the emphasis given in the previous section
that 
%the effect of inclination is greatly removed with the $C_e$ parameter. 
using $C_e$ parameter removed the effect of inclination.
The reason is that early and late spiral galaxies are
not well separated in the concentration parameter space, 
%and show large dispersion that heavily overlaps.
and show large dispersion with heavy overlaps.
So the effect of inclination
does not play a crucial role. In fact, visual separation into early and
late spirals relies more on the opening of spiral arms and texture.

It is important to note that contaminants from late spirals to the early-type
sample, or vice versa, are very small, less than $\la 1$\%, at least 
with the $C_e$ and $Y$ indices. 
Most of the contaminants in the E+S0 sample are from
S0a and Sa galaxies.

%Finally, we examine how the performance improves with the use of   
%the $C_e$-$Y$ correlation. 
Finally, we examine how the performance of our morphology classifier
improves by considering 2-dimensional classification in the $C_e$-$Y$
space.
The result is shown in 
Table \ref{3x3tbl_2}. The completeness of the Sa+Sb galaxy sample
is $314/461=68.1$\%, a 2\% increase, 
compared with the value found employing $Y$ alone. 
The contamination in the $\ell_{\rm I}$ sample 
from E+S0 galaxies and late-type spiral galaxies
decreases to $(53+137)/504=34.1$\%, which is 5\% smaller 
than the value obtained by using $Y$ alone.
The contamination in the early-type spiral sample still primarily arises
from late-type spiral galaxies, rather than E+S0 galaxies. 
The contamination in the $\ell_{\rm II}$ sample
decreases from 17.9\% to 17.2\%.

\section{CONCLUSIONS}
\label{conclusions}

%We attempted to find photometric parameters that correlate well with
%the visually-classified morphology of galaxies. 
%We first examined the standard concentration index $C=r_{50}/r_{90}$ 
%defined with the Petrosian flux in the circular
%apertures 
%and found that the correlation is significantly affected by inclination.
%The standard concentration index $C$ of highly
%inclined spiral becomes smaller value
%as if the galaxy has more centrally concentrated light
%of earlier types.
%We found that this effect is removed by 
%defining 
We started by examining the standard concentration index $C=r_{50}/r_{90}$,
defined using the Petrosian flux in circular apertures, and found that
the correlation is significantly affected by galaxy inclination. The
value of the standard concentration index $C$ of a highly inclined
spiral is artificially reduced (i.e. the galaxy appears to have more
centrally concentrated light) and the galaxy is misclassified as an
early type. We found that this inclination dependence vanishes if we
define 
the concentration index using elliptical apertures.
The ellipse-based concentration index $C_e$ calculated from the Petrosian flux
gives an inclination-independent indicator.
%Thus, the correlation of the concentration index with  $T$ is improved.

In addition, we devised a new texture parameter, $Y$, that represents
the coarseness of surface brightness. 
%The $Y$ parameter measures the magnitude of texture of galaxies 
%seen in the disk, compared with the overall flux contrast
%including the bulge component. 
%The quantity $Y$ is defined
%in a manner that works rather closely to the way we 
%carry out visual classification. 
%The $Y$ parameter acquires large values for 
%galaxies that have distinct arms 
%or those that do not show conspicuous texture, but have
%inconspicuous bulges, such as those of the Magellanic Clouds.
%We found a good correlation of the $Y$ parameter with 
%visually obtained morphology.
The $Y$ parameter measures the texture of a galaxy disk in relation to
the galaxy's overall surface brightness contrast (including the galaxy
bulge). It is defined in a manner that closely mimics visual
classification. Late type spiral galaxies (which often show distinct
spiral arms) or Magellanic type irregulars (whose disks are less
pronounced but so are their bulges) both have large $Y$ values, in
contrast to early type galaxies. We found that the $Y$ parameter of a
galaxy is strongly correlated with its visual morphology.

%The performance of 
%$C$, $C_e$ and $Y$ as classifiers for early--late morphological types and
%for three types (early type, early spirals and late spirals) was investigated.
%We found superior performance of  $C_e$ over the standard
%$C$, but even better classification was attained with the
%$Y$ parameter. For example, we could achieve 
%a degree of completeness as 
%high as 88\%, or of contamination as low as 12\%.
%
We investigated the performance of the three different photometric
parameters ($C$, $C_e$, and $Y$) for morphological classification into two
(E+S0 and Sa+Sb+Sc+Sdm+Im) or three (E+S0, Sa+Sb, and Sc+Sdm+Im)
galaxy types. In both cases we found that the elliptical aperture
classifier, $C_e$, is better than the standard circular aperture
classifier, $C$. The coarseness parameter, $Y$, produces results superior
to those obtained with either of the concentration index parameters.
Depending on the desired balance between completeness and
contamination, sample completeness as high as 88\% or contamination as
low as 12\% is achievable.

%We also tried to find the ``best classifier'' in the two-parameter space of
%$C_e$ and $Y$. The highest success rate is 
%89\% for completeness, or 12\% contamination for 
%early--late classification. We attain 
%68\% completeness for three types classification.
We can further improve the classification by considering
2-dimensional (using the $C_e$-$Y$ plane) classification. In the case of
classification into two morphological types, this allows us to attain
a 89\% completeness and contamination as low as 12\%. For classification
into three morphological types, the completeness is 68\%.

%It is important that a photometric parameter which represents 
%galaxy morphology quite well is newly devised in the epoch of large CCD
%surveys. The parameter, coarseness, has a 
%Our new texture parameter may become a clue which automates
%morphological classification of galaxies.
Our newly devised photometric parameter, coarseness, provides 
%such a good
a mode of
morphological classification as good as
the traditionally used human-eye
classification. At the same time, it is fully automated, and thus, can
be used quite easily for millions of galaxies. Therefore, the coarseness
parameter presented in this paper has a potential to open new doors to
%detailed studies of galaxy morphology in future large CCD surveys.
detailed studies of galaxy morphology in current and future large CCD surveys.\\

%In view of the fact that visual classification
%contains subjective elements of uncertainty, this seems to 
%be perhaps nearly the best one could achieve.

%{
%\parindent=0pt
%{\bf ACKNOWLEDGMENTS}
%}
\acknowledgments

CY thanks Satoru Ikeuchi, Takahiko Matsubara and Tomoyuki Hanawa
for useful discussions.
We thank David Bazell for his reading of the coarseness parameter
description and helpful comments.
We acknowledge Arunas Kucinskas and Kiyotaka Tanikawa
for their extensive help in English issues. 
We thank Linux, XFree86, and other UNIX-related communities
for development of various useful software.
This research has made use of the Plamo Linux.\\
~\\

%This paper is based on the master's thesis of one of the authors(C.Y.).

%\clearpage

\small
\baselineskip=10pt

%%%%%%%%%%%%%%%%%%%%%%%%%%%%%%%%%%%%%%%%%%%%%%%%%%%%%%%%

%%%%%%%%%%%%%%%%%%%%%%%%%%%%%%%%%%%%%%%%%%%%%%%%%%%%%%%%

\end{document}